# Phase-Singularity Control of Topological Acoustic Spin Textures

Huaijin Ma[1†], Te Liu[1†], Jiachen Sheng[1], Xiaochang Pan[1], Wenwei Qian[1],
Xiangyu Chen[1], Kaiyuan Cao[1*], Jinpeng Yang[1*], Jian Wang[1*]

*1. College of Physical Science and Technology, Yangzhou University, Yangzhou 225002, China*

**Abstract:**

Acoustic spin textures provide a degree of freedom for programmable topological field states, yet a unified framework for constructing and classifying higher-order textures in open-air acoustics remains lacking. Here we introduce polygonal standing-wave interference as a framework connecting discrete symmetry, angular phase encoding, phase singularities, chiral time-averaged energy flux, and acoustic spin. By controlling the number and relative phases of the standing-wave channels, we experimentally realize $D_4$ meron/anti-meron lattice, $D_6$ double-skyrmion superlattice, and $D_8$ acoustic-spin quasicrystal composed of four meron/anti-meron sublattices. Finite polygonal fields are normalized angular discretizations of continuous standing-wave spectra. Under the $m = 2$ phase encoding, the $D_6$ and $D_8$ fields reproduce the local $q = +2$ core of a second-order Bessel vortex while retaining distinct long-range order through their discrete interference channels. At fixed $m$ increasing $N$ under normalized angular sampling yields the continuous Bessel limit. These results establish phase encoding, pairwise interference, reciprocal-space composition, and angular discretization as a unified design principle for programmable topological acoustic spin textures.

† These authors contributed equally to this work.

* Corresponding author: Kaiyuan Cao (kycao@yzu.edu.cn),
Jinpeng Yang (yangjp@yzu.edu.cn),
Jian Wang (wangjian@yzu.edu.cn).

***Introduction—***Topological textures [1–3] are vector-field configurations characterized by nontrivial mappings and classified by topological numbers. Combining tunable length scales, low driving thresholds, and topological protection against defects and perturbations [4,5], they are regarded as potentially transformative functional units for next-generation information and energy technologies [6]. Representative examples, including skyrmions [7,8] and merons/anti-merons [9], can be described as continuous vector mappings from real-space [9] or momentum-space [10] manifolds onto an order-parameter sphere and classified by the homotopy group $\pi_m\left(S^n\right)$, which characterizes a unit-vector field with $n$ degrees of freedom defined on an $m$-dimensional manifold [11,12]. Experimentally, such nontrivial mappings in magnetic systems [13–15] commonly emerge from Dzyaloshinskii-Moriya interactions or from competition between frustrated symmetric exchange and external fields, forming localized topological magnetic domains encoded in the magnetic-moment order parameter. In wave systems [6,9,16–19], by contrast, topological textures originate from coherent coupling among continuous field components and their spatial organization, manifesting as classical-field topological quasiparticles with well-defined topological charges and collective dynamical behavior. Existing construction schemes, however, generally rely on specific interactions, structural constraints, or field superpositions, while a general route and dynamical mechanism for constructing topological textures across physical systems remain lacking.

In wave systems, elucidating how wave-field phase topology induces vector topological textures is central to understanding their formation mechanism. Here we take phase singularities as the starting point to trace the underlying physical process through which topological textures emerge. Phase singularities are fundamental carriers of wave-field phase topology, appearing as local defects at which the complex-wave amplitude vanishes and the phase becomes undefined [20]. Along any closed contour $C$ enclosing an isolated singularity, the accumulated phase satisfies $2\pi q$, where the integer $q$ is the topological charge conserved under continuous perturbations. Such singularities are ubiquitous in helical beams carrying orbital angular momentum and in complex wave systems [21,22]. They are typically associated with rapid or discontinuous phase variations, causing the wavefront and the corresponding time-averaged energy flux (Poynting vector) to circulate around the singularity, while the resulting nonzero local vorticity is intimately connected to the spin vector field [23]. Accordingly, control of phase singularities has evolved from manipulating individual vortices and assigning their topological charges to globally engineering the spatial organization and propagation dynamics of multiple-singularity systems [24–26]. Meanwhile, singularity engineering has been extended into the spatiotemporal domain and along the propagation dimension, producing spatiotemporal vortices carrying transverse orbital angular momentum [27] and enabling deterministic longitudinal evolution of topological charge through singularity migration [28], splitting, and annihilation [29]. These processes can further induce transverse angular-momentum transport, additional geometric phases, and distance-dependent dispersion and torque responses, providing new controllable degrees of freedom for high-dimensional information transmission, precision measurement, and wave-matter interactions. Although these advances have established singularities as designable topological units and a central element of wave-field engineering, the mechanisms governing the formation and evolution of spatial topological-texture configurations from phase singularities remain unresolved.

Taking advantage of the long wavelengths and direct observability of acoustic fields [30–35], here we use acoustics to investigate the dynamical mechanism by which phase singularities induce topological vector textures. Through a polygonal standing-wave acoustic-field model, we establish a unified framework for constructing acoustic-spin topological textures from phase singularities. We reveal the fundamental connection among phase singularities, chiral energy flux, and acoustic spin, together with the mechanism underlying the generation of the acoustic-spin vector: a topological charge $q=+1$ ( $q=-1$ ) induces right-handed (left-handed) time-averaged energy flux and locks the positive (negative) polarity of the acoustic-spin component. More generally, for $N$ nonparallel standing-wave channel orientations, we derive general expressions for the acoustic pressure, time-averaged energy flux, and acoustic spin, demonstrating that the interchannel phase differences select the energy-flux chirality, the relative channel orientations determine the geometric interchannel coupling strength, and the standing-wave position factors control the spin polarity and nodal distribution. Based on this mechanism, increasing the number of nonparallel standing-wave channels produces a meron/anti-meron lattice, a double-skyrmion superlattice, and an acoustic-spin quasicrystal. Using the wave-vector coupling relations between neighboring channels, we further confirm the superposition dynamics underlying the formation of the spin superlattice and quasicrystal. As the discrete polygonal angular spectrum approaches the continuous limit $N \to \infty$, the acoustic field further converges to a Bessel field, which not only explains the second-order phase singularities observed in the $D_6$ and $D_8$ configurations but also reveals the limiting approximation of the polygonal standing-wave acoustic-field model. This work advances singularity engineering from shaping singularity geometries to the predictable generation of vector topological states and provides a transferable design principle for programmable spin topology in structured waves.

***Phase-singularity framework.***—A phase singularity is a fundamental topological defect of a complex wave field. For a complex acoustic pressure $p(\mathbf{r}) = P_0(\mathbf{r})e^{i\phi(\mathbf{r})}$, it occurs at an isolated amplitude zero, $P_0 = 0$, where the phase $\phi$ is undefined [36]. The phase winding around the singularity is quantified by the integer topological charge

$$q = \frac{1}{2\pi}\oint_C \nabla\phi(\mathbf{r}) \cdot d\mathbf{l}, \tag{1}$$

where $C$ is a closed contour enclosing one isolated singularity, crosses no other singularity, and satisfies $P_0(\mathbf{r}) \neq 0$ along the contour [Fig. 1(a)]. The charge constrains the local phase winding and the circulation of phase-derived observables, thereby connecting wavefront dislocations, vortex waves, and classical-field topological textures.

In acoustics, this topological phase structure directly governs energy transport. The linearized Euler equation gives the time-averaged energy flux associated with $p(\mathbf{r}) = P_0(\mathbf{r})e^{i\phi(\mathbf{r})}$ as

$$\mathbf{J}(\mathbf{r}) = \frac{1}{2\omega\rho_0} P_0^2(\mathbf{r}) \nabla\phi(\mathbf{r}). \tag{2}$$

where $\rho_0$ is the mass density of air and $\omega = 2\pi f$ is the angular frequency; the derivation is given in Sec. S1 of the Supplemental Material [36]. Equation (2) shows that the phase gradient directs the local energy flow, whereas $P_0^2(\mathbf{r})$ weights its magnitude. Phase winding around a singularity therefore generates a local chiral circulation [20,31]. For the conjugate $q = \pm 1$ cores considered here, reversing the phase winding reverses the local circulation, although the detailed energy flux pattern also depends on the amplitude and interference geometry [Fig. 1(b)].

Within the same assumptions, the acoustic spin density for a monochromatic field with acoustic velocity $\mathbf{u}$ is [23,33]

$$\mathbf{S} = \frac{\rho_0}{2\omega}\operatorname{Im}\left(\mathbf{u}^* \times \mathbf{u}\right) = \frac{1}{\omega^2}\nabla \times \mathbf{J}, \tag{3}$$

Equation (3) relates a nonzero curl of the time-averaged energy flux to the out-of-plane spin in a two-dimensional field [Fig. 1(c)]. In the present spoof surface acoustic wave (SSAW) implementation [6,32], evanescent confinement supplies the in-plane spin components needed to form the measured three-dimensional texture. Phase winding organizes the chiral flux, while amplitude weighting, pairwise interference, and confinement determine the complete spin texture.

Extending this local mechanism to programmable real-space textures requires simultaneous control of the positions and charges of phase singularities and multiple nonparallel interference channels. We begin with a single standing-wave channel, constructed by coherently superposing two equal-amplitude SSAW components propagating oppositely along the metasurface [Fig. 1(d)]. The two counterpropagating constituents share the same harmonic phase $\varphi_j$, and their complex pressures are

$$p_j^{\pm}\left(\mathbf{r}_{\parallel}, z\right) = A_j e^{i\varphi_j} e^{\pm ik\hat{\mathbf{n}}_j \cdot \mathbf{r}_{\parallel}} e^{-\tau z}, \tag{4}$$

where $\mathbf{r}_{\parallel} = (x, y)$. Their superposition gives

$$p_j(\mathbf{r}_{\parallel},z) = p_j^+ + p_j^- = 2A_j e^{i\varphi_j}\cos[k\hat{\mathbf{n}}_j\cdot\mathbf{r}_{\parallel}]e^{-\tau z}. \tag{5}$$

An ideal standing-wave channel therefore has no net propagation direction and is instead characterized by the orientation of its in-plane axis. We represent the axis of channel $j$ by the unit vector

$$\hat{\mathbf{n}}_j = (\cos\Theta_j, \sin\Theta_j), \quad j = 0,1,...,N-1, \tag{6}$$

where $\Theta_j = j\pi/N$ specifies the channel-axis orientation relative to the reference $x$ axis. Polygonal standing-wave interference then provides a natural multichannel realization [Fig. 1(d)]. The $2N$ boundary-launched SSAW components form $N$ equal-phase counterpropagating pairs and hence $N$ independent standing-wave channels. Summing these channels gives the total complex acoustic field

$$p_N(\mathbf{r}_{\parallel},z) = 2e^{-\tau z}\sum_{j=0}^{N-1} A_j e^{i\varphi_j}\cos[k\hat{\mathbf{n}}_j\cdot\mathbf{r}_{\parallel}]. \tag{7}$$

Here $A_j$ is the amplitude of each constituent SSAW. The $D_4$, $D_6$, and $D_8$ fields considered below correspond to $N$=2, 3, and 4. Substitution of Eq. (7) into Eqs. (2) and (3) yields pairwise-interference rules for the in-plane energy flux and out-of-plane spin. Omitting positive prefactors that do not affect handedness, polarity, or topology, we obtain

$$\mathbf{J}_{N,\parallel} \propto \sum_{j<l} A_j A_l \sin\Delta\varphi_{jl}\left(\cos u_l \sin u_j \hat{\mathbf{n}}_j - \cos u_j \sin u_l \hat{\mathbf{n}}_l\right), \tag{8}$$

and

$$\mathbf{S}_{N,z} \propto \sum_{j<l} A_j A_l \sin\Delta\varphi_{jl}\sin\Delta\Theta_{jl}\sin u_l \sin u_j. \tag{9}$$

where $\Delta\varphi_{jl} = \varphi_l - \varphi_j$, $\Delta\Theta_{jl} = \Theta_l - \Theta_j$, and $u_j = k\hat{\mathbf{n}}_j\cdot\mathbf{r}$. Full derivations of Eqs. (8) and (9) are given in Sec. S2 of the Supplemental Material [36]. Equation (8) shows that the in-plane time-averaged energy flux arises entirely from interference between distinct standing-wave channels. The factor $\sin\Delta\varphi_{jl}$ is the interchannel phase factor, for $\Delta\varphi_{jl} = 0$ or $\pi$, the pair contributes no active flux, whereas a nontrivial interchannel phase difference generates a directional pairwise contribution. Reversing $\Delta\varphi_{jl}$ reverses the contribution of that pair. With the other factors fixed, a larger magnitude of $|\sin\Delta\varphi_{jl}|$ produces a stronger chiral-flux contribution.

Equation (9) gives the design rule for polygonal acoustic spin textures. The factor $\sin\Delta\varphi_{jl}$ selects the active channel pairs and fixes the sign of each contribution. The geometric factor $\sin\Delta\Theta_{jl}$ quantifies the relative angular coupling between channel orientations, whereas the spatial factor sets the local polarity, nodes, and modulation. Increasing $N$ introduces additional pairwise-interference channels that, under suitable phase encoding, can generate multiscale or quasiperiodic textures.

To characterize the spatial scales introduced by different interference channels, we define the sum and difference wave vectors

$$\begin{aligned}\mathbf{K}_{jl}^{+} &= k(\hat{\mathbf{n}}_j + \hat{\mathbf{n}}_l),\\ \mathbf{K}_{jl}^{-} &= k(\hat{\mathbf{n}}_j - \hat{\mathbf{n}}_l).\end{aligned} \tag{10}$$

Since $u_j = k\hat{\mathbf{n}}_j\cdot\mathbf{r}_{\parallel}$ and $u_l = k\hat{\mathbf{n}}_l\cdot\mathbf{r}_{\parallel}$, the spatial factor in Eq.(9) can be recast as

$$\sin u_l \sin u_j = \frac{1}{2}\left[\cos\left(\mathbf{K}_{jl}^{-}\cdot\mathbf{r}_{\parallel}\right) - \cos\left(\mathbf{K}_{jl}^{+}\cdot\mathbf{r}_{\parallel}\right)\right]. \tag{11}$$

Thus, each pairwise contribution to $S_z$ contains both difference- and sum-wave-vector harmonics. Using Eq. (6), their magnitudes are

$$\left|\mathbf{K}_{jl}^{+}\right| = 2k\cos\left(\frac{s\pi}{2N}\right),$$
$$\left|\mathbf{K}_{jl}^{-}\right| = 2k\sin\left(\frac{s\pi}{2N}\right). \tag{12}$$

where $s = l - j = 1,2,...,N-1$ labels the angular separation of the interfering channels. Different pairwise channels can populate common reciprocal-space shells characterized by the same $|\mathbf{K}|$. Increasing $N$ introduces additional pairwise-interference channels and spatial harmonics, enabling the spin texture to evolve from a single-scale lattice to multiscale and quasiperiodic structures.

For small $N$, the finite polygonal field defined by Eqs. (6) and (7) can also be viewed as a discrete sample of a continuous standing-wave angular spectrum. In the limit $N \to \infty$, equal normalized amplitudes yield the continuous Bessel spectrum [37]. With angular phase encoding $\varphi_j = m\Theta_j$, the in-phase choice $m = 0$ approaches the zeroth-order Bessel field $p_\infty \propto J_0(kr)$, whereas the simplest nontrivial standing-wave encoding, $m = 2$, produces the central charge $q = 2$ of the second-order Bessel vortex $p_\infty \propto J_2(kr)e^{i2\alpha}$; here $\alpha$ is the polar angle. Finite polygonal fields are thus discrete realizations of continuous Bessel spectra with different angular sampling and phase encoding. In this framework, $N = 2$ is the lowest-order $D_4$ configuration. We use it next to establish the basic mapping among phase singularities, chiral energy flux, and a meron/anti-meron lattice.

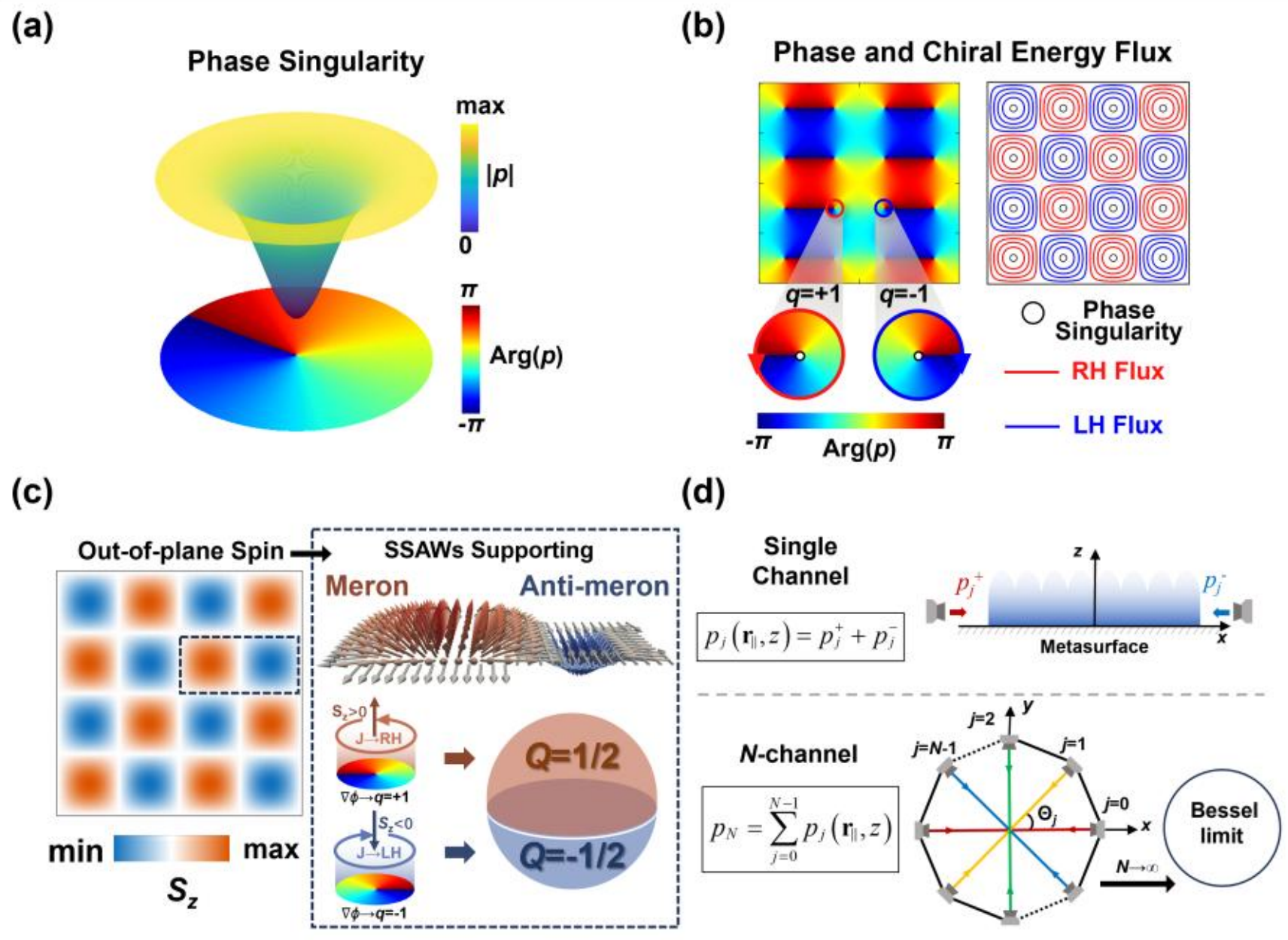

**FIG. 1. Phase-singularity framework for polygonal acoustic-spin topology.** (a) Pressure amplitude $|p|$ and phase arg($p$) near an isolated singularity; the phase winding around a closed contour defines $q$. (b) Opposite charges $q$=±1 and the corresponding right- and left-handed time-averaged energy-flux circulations. (c) Alternating out-of-plane spin $S_z$ generated by the flux curl and the SSAW-supported meron/anti-meron textures, with $Q$=±1/2. (d) A single standing-wave channel is formed by an equal-phase pair of counterpropagating SSAWs; 2$N$-sided excitation provides $N$ standing-wave channels; under normalized sampling, the angular spectrum becomes continuous as $N$→∞.

***$D_4$ meron/anti-meron lattice.***—We experimentally realize the minimal $D_4$ field using two equal-amplitude orthogonal standing-wave channels. Choosing channel 0 as the phase reference $\varphi_0 = 0$, we set the orthogonal channel to $\varphi_1 = \pi/2$, so that the interchannel phase difference is $\Delta\varphi_{01} = \pi/2$ [Fig. 2(a)]. This encoding gives $|\sin\Delta\varphi_{01}| = 1$, while the orthogonal geometry gives $|\sin\Delta\Theta_{01}| = 1$, according to Eqs. (8) and (9), the phase and geometric factors therefore jointly maximize the chiral energy-flux and out-of-plane-spin contributions of the single interference channel. Experimental details are provided in Sec. S3 of the Supplemental Material [36].

Both simulated and measured complex-pressure fields exhibit periodic near-zero-amplitude nodes and a checkerboard phase pattern [Figs. 2(b) and 2(c)]. The phase-singularity positions satisfy

$$\left(x_\mu, y_\nu\right) = \left(\frac{\frac{\pi}{2} + \mu\pi}{k_{ssaw}}, \frac{\frac{\pi}{2} + \nu\pi}{k_{ssaw}}\right), \qquad \mu, \nu \in \mathbb{Z}. \tag{13}$$

in agreement with the minimal model $D_4$. Closed-loop phase integration around two neighboring singularities gives windings of $+2\pi$ and $-2\pi$, corresponding to charges $q_1 = +1$ and $q_2 = -1$, respectively [Fig. 2(d)]. The experiment therefore realizes an alternating lattice of phase singularities. The orthogonal pressure field and its phase-singularity conditions are derived in Secs. S4 of the Supplemental Material [36].

We next reconstruct the three-dimensional particle-velocity field from the complex pressure and calculate the acoustic spin using Eq. (3). Simulation and experiment show that the alternating singularities induce alternating out-of-plane spin polarities. Evanescent SSAW confinement supplies the in-plane components, producing local meron/anti-meron textures [Fig. 2(e)]. Because the texture is defined by the complex field rather than by the instantaneous velocity, it is stationary in time; the oscillatory instantaneous-velocity texture in the $\varphi_1 = 0$ -field limit is discussed in Sec. S5 of the Supplemental Material [36]. For the minimal $D_4$ model, the field depends directly on the channel amplitudes and the interchannel phase difference $\Delta\varphi_{10}$, providing simple control of the spin polarity and magnitude. The corresponding tuning is demonstrated in Sec. S6 of the Supplemental Material [36] for the $D_4$ configuration.

For a quantitative topological characterization, we calculate the skyrmion density[6]

$$\rho_Q = \frac{1}{4\pi}\mathbf{n}\cdot\left(\partial_x\mathbf{n}\times\partial_y\mathbf{n}\right), \qquad \mathbf{n} = \frac{\mathbf{S}}{|\mathbf{S}|}. \tag{14}$$

The $\rho_Q$ distribution has the same alternating periodicity as the spin texture [Fig. 2(f)]. Integrating over the central dashed region gives $Q = -0.0006$ in simulation and $Q = -0.0064$ in experiment, both close to zero because the meron and anti-meron contributions cancel within that region. Tests with structural and excitation defects further show that the alternating texture remains identifiable under the perturbations considered (Sec. S6 of the Supplemental Material [36]). These results demonstrate that the $D_4$ polygonal standing-wave field generates a spatially periodic, time-stationary acoustic-spin meron/anti-meron lattice.

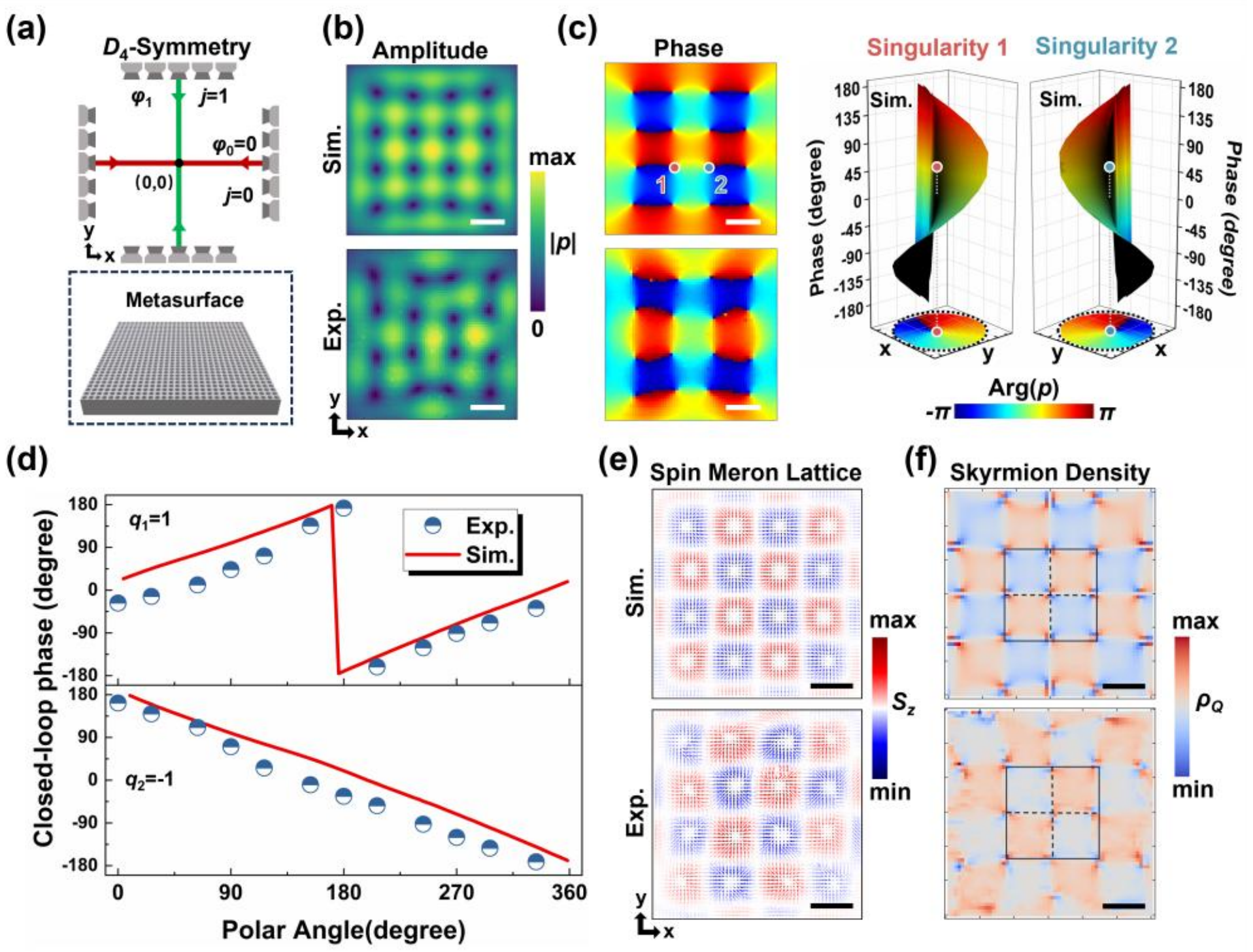

**FIG. 2. $D_4$ meron/anti-meron lattice.** (a) $D_4$ boundary excitation and the perforated acoustic metasurface. (b), (c) Simulated (top) and measured (bottom) pressure amplitude $|p|$ and phase $\arg(p)$, the local phase surfaces in (c) identify neighboring singularities 1 and 2 (The scale bar is 3 cm). (d) Simulated (lines) and measured (symbols) phase winding along closed loops around the two singularities, yielding $q_1$=+1 and $q_2$=−1. (e) Simulated (top) and reconstructed (bottom) spin textures; color denotes $S_z$ and arrows show the in-plane components (The scale bar is 3 cm). (f) Corresponding skyrmion-density maps; the boxes mark the integration regions (The scale bar is 3 cm).

***$D_6$ double-skyrmion superlattice.***—In the minimal $D_4$ model, neighboring $q = \pm 1$ phase singularities generate alternating chiral flux and a meron/anti-meron lattice. To test the generality of this mechanism at higher polygonal order, we extend the excitation to $D_6$ symmetry. The hexagonal source contains three standing-wave channels oriented at $\hat{\mathbf{n}}_0$, $\hat{\mathbf{n}}_1$, and $\hat{\mathbf{n}}_2$. For equal amplitudes, the topology is governed primarily by the interchannel phase relations. We choose channel 0 as the phase reference $\varphi_0 = 0$, so that the two independent phase offsets $(\varphi_1, \varphi_2)$ span a two-dimensional relative-phase parameter space [Fig. 3(a)]. To identify an analytically tractable nontrivial phase structure, we examine the complex pressure at the center, $\mathbf{r} = 0$. The parameter space contains two isolated zeros:

$$(\varphi_1, \varphi_2) = \left(\frac{2\pi}{3}, \frac{4\pi}{3}\right) \text{ and } \left(\frac{4\pi}{3}, \frac{2\pi}{3}\right). \tag{15}$$

The two zeros have opposite phase windings and therefore select phase encodings of opposite handedness. We choose $(\varphi_1, \varphi_2) = \left(\frac{2\pi}{3}, \frac{4\pi}{3}\right)$ as the $D_6$ encoding; alternative encodings are discussed in Sec. S7 of the Supplemental Material [36]. This choice cancels the zeroth-order pressure at the center. Because the standing-wave expansion contains no linear term about the origin, the leading nonzero contribution is quadratic:

$$p_3(r, \phi) \propto r^2 e^{i2\phi}. \tag{16}$$

The phase therefore accumulates by $4\pi$ along a loop around the center, giving the second-order singularity $q = +2$. The opposite parameter-space zero, $(\varphi_1, \varphi_2) = \left(\frac{4\pi}{3}, \frac{2\pi}{3}\right)$, yields a high-order core with $q = -2$. This encoding may also be interpreted as an effective orbital-angular-momentum (OAM) charge $l = 2$ under sixfold discrete symmetry, whose local phase structure is a second-order topological core [24].

Simulated and measured pressure fields verify this phase-selection mechanism. The amplitude shows a near-zero central core with multiscale peripheral modulation [Fig. 3(b)], while the phase reveals a second-order central singularity surrounded by a network of first-order singularities [Fig. 3(c)]. Closed-loop integration of the local phase surface gives a winding of approximately $4\pi$ in both simulation and experiment, confirming the $q = 2$ high-order core [Fig. 3(d)].

This high-order phase structure produces a hierarchical acoustic-spin superlattice. In the spin texture reconstructed from the complex pressure, the sign of $S_z$ remains controlled by the local phase singularities, while its magnitude, nodes, and polarity pattern are modulated by a longer-scale envelope [Fig. 3(e)]. Horizontal and vertical centerline profiles show agreement between experiment and simulation in the polarity domains, local zeros, and modulation periods, establishing an observable spin superlattice in the $D_6$ standing-wave field. A pronounced spin defect occurs at the central $q = +2$ phase core, indicating that a high-order singularity can act as a local organizing center for the spin texture. This behavior is qualitatively consistent with that reported for an $l = 2$ plasmonic-vortex field [24].

According to Eq. (12), the multiscale structure of the $D_6$ field originates from the sum- and difference-wave-vector harmonics of the three pairwise interference channels. For the selected $D_6$ state, $\Theta_j = j\pi/3, j = 0,1,2$. The pairs (0,1) and (1,2) have $s$=1, whereas (0,2) has $s$=2. The $s$=1 and $s$=2 channels interchange the sum- and difference-wave-vector magnitudes while collectively organizing the resulting harmonics into two reciprocal-space shells,

$$\begin{aligned} &\left\{\mathbf{K}_{01}^-, \mathbf{K}_{12}^-, \mathbf{K}_{02}^+\right\}, \quad |\mathbf{K}| = K_< = k, \\ &\left\{\mathbf{K}_{01}^+, \mathbf{K}_{12}^+, \mathbf{K}_{02}^-\right\}, \quad |\mathbf{K}| = K_> = \sqrt{3}k. \end{aligned} \tag{17}$$

Here $K_<$ and $K_>$ denote the lower- and higher-magnitude reciprocal-space shells; they are not labels for the individual sum or difference wave vectors $\mathbf{K}_{jl}^{\pm}$. Grouping the corresponding harmonics by shell, the in-plane energy flux and out-of-plane spin decompose as

$$\mathbf{J}_{3,\|} = \mathbf{J}_{3,\|}^{(<)} + \mathbf{J}_{3,\|}^{(>)}, \qquad S_{3,z} = S_{3,z}^{(<)} + S_{3,z}^{(>)}. \tag{18}$$

The analytical decomposition in Sec. S7 of the Supplemental Material [36] shows that the $K_<$ and $K_>$ shells generate circulation networks and out-of-plane-spin subtextures with distinct spatial scales [Figs. 3(f) and 3(g)]. Their superposition yields the double-skyrmion superlattice comprising two interpenetrating skyrmion sublattices [Fig. 3(e)]. Thus, the superlattice follows directly from the organization of pairwise sum- and difference-wave-vector harmonics into two characteristic reciprocal-space shells.

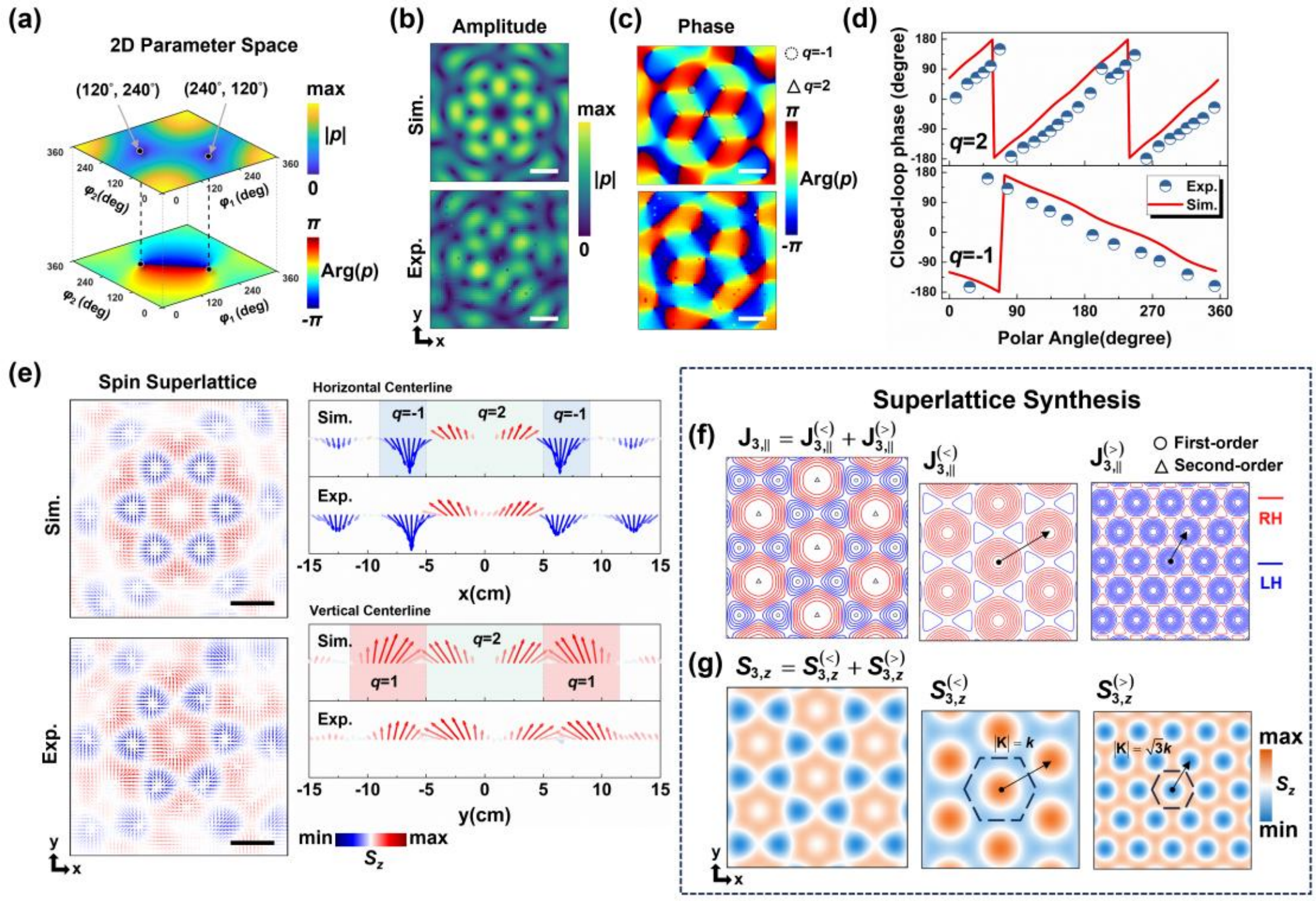

**FIG. 3. Parameter-space-selected $D_6$ double-skyrmion superlattice.** (a) Pressure amplitude $|p|$ and phase arg($p$) in the ($\varphi_1$,$\varphi_2$) parameter space; the two isolated zeros correspond to opposite phase encodings. (b), (c) Simulated (top) and measured (bottom) pressure amplitude and phase; symbols in (c) mark the central ($q$=+2) singularity and surrounding ($q$=-1) singularities (scale bars, 3 cm). (d) Simulated (lines) and measured (symbols) phase winding along closed loops enclosing representative ($q$=+2) and ($q$=-1) singularities. (e) Simulated and reconstructed spin superlattices with horizontal and vertical centerline profiles; color denotes $S_z$, and arrows show the in-plane spin components (scale bars, 3 cm). (f), (g) Shell-resolved in-plane energy flux and out-of-plane spin for $|\mathbf{K}| = K_< = k$ and $|\mathbf{K}| = K_> = \sqrt{3}k$.

***$D_8$ acoustic spin quasicrystal.***—In the $D_6$ system, parameter-space selection among three standing waves produces a high-order phase core and a double-skyrmion superlattice. To add angular interference channels beyond the constraints of a two-dimensional translational lattice, we extend the field to $D_8$ symmetry. The octagonal source comprises four standing-wave channels oriented at $\hat{\mathbf{n}}_0$, $\hat{\mathbf{n}}_1$, $\hat{\mathbf{n}}_2$, and $\hat{\mathbf{n}}_3$. Compared with the $D_4$ and $D_6$ cases, the $D_8$ system contains more noncollinear pairwise-interference channels and can therefore support an aperiodic spin texture with long-range rotational order.

For equal amplitudes, we choose channel 0 as the phase reference $\varphi_0 = 0$, leaving three independent phase offsets $(\varphi_1, \varphi_2, \varphi_3)$ that span the three-dimensional relative-phase parameter space. Unlike the isolated zeros of the $D_6$ case, zeros of the central complex pressure form singular lines in this parameter space [Fig. 4(a)]. Their complete parameterization is given in Sec. S7 of the Supplemental Material [36]. A generic point on a singular line guarantees cancellation of only the zeroth-order central field. We therefore select the encoding

$$(\varphi_1, \varphi_2, \varphi_3) = \left(\frac{\pi}{2}, \pi, \frac{3\pi}{2}\right), \tag{19}$$

This encoding cancels the zeroth-order central pressure while retaining an angular phase relation compatible with the $D_8$ excitation geometry. The pairs (0,1), (1,2), (2,3), (0,3) are active, with $\sin\Delta\varphi_{jl} = +1, +1, +1, -1$, respectively. The (0,2) and (1,3) pairs instead have $\Delta\varphi_{jl} = \pi$, so $\sin\Delta\varphi_{jl} = 0$ and they contribute neither $\mathbf{J}_{4,\parallel}$ nor $S_{4,z}$. Simulated and measured pressure amplitudes show eightfold modulation and multiscale local nodes [Fig. 4(b)], while the phase contains a network of first- and second-order singularities [Fig. 4(c)]. The experiment reconstructed and simulated spin textures agree in their polarity pattern and local topological units [Fig. 4(d)].

To assess quasiperiodic order, we use the analytical solution to examine the large-area $S_z$ distribution and its reciprocal-space spectrum. The real-space field has no common two-dimensional translational unit cell, while its FFT contains eightfold-related peaks [Fig. 4(e)]. The combination of rotational order and the absence of a common translation support the identification of the texture as quasiperiodically ordered. According to Eqs.(11) and (12) this quasiperiodic structure originates from the sum- and difference-wave-vector harmonics of the active pairwise interference channels. For the selected $D_8$ state,

$$\Theta_j = \frac{j\pi}{4}, \qquad \varphi_j = \frac{j\pi}{2}, \qquad j = 0,1,2,3. \tag{20}$$

The pairs $(0,1)$, $(1,2)$, and $(2,3)$ have $s=1$, whereas $(0,3)$ has $s=3$. The remaining pairs $(0,2)$ and $(1,3)$, are inactive because $\sin\Delta\varphi_{jl} = 0$. The $s=1$ and $s=3$ channels interchange the sum- and difference-wave-vector magnitudes, but the nonzero harmonics populate only two reciprocal-space shells,

$$\begin{aligned} &\left\{\mathbf{K}_{01}^{-}, \mathbf{K}_{12}^{-}, \mathbf{K}_{23}^{-}, \mathbf{K}_{03}^{+}\right\}, \quad |\mathbf{K}| = K_{<} = 2k\sin\frac{\pi}{8}, \\ &\left\{\mathbf{K}_{01}^{+}, \mathbf{K}_{12}^{+}, \mathbf{K}_{23}^{+}, \mathbf{K}_{03}^{-}\right\}, \quad |\mathbf{K}| = K_{>} = 2k\cos\frac{\pi}{8}. \end{aligned} \tag{21}$$

Unlike the $D_6$ case, each shell contains four wave-vector directions that separate into two mutually rotated sublattice sets. The out-of-plane spin can therefore be decomposed as

$$S_{4,z} = S_{4,zA}^{(<)} + S_{4,zB}^{(<)} + S_{4,zA}^{(>)} + S_{4,zB}^{(>)}. \tag{22}$$

Here $S_{4,zA}^{(<)}$ and $S_{4,zB}^{(<)}$ are two long-period meron/anti-meron sublattices associated with the $K_{<}$ shell and mutually rotated by $\pi/4$, whereas $S_{4,zA}^{(>)}$ and $S_{4,zB}^{(>)}$ are two short-period meron/anti-meron sublattices associated with the $K_{>}$ shell. Their superposition produces the total $S_{4,z}$ field with eightfold rotational order but no simple translational period. The $D_8$ texture is therefore an acoustic spin quasicrystal composed of four distinguishable meron/anti-meron sublattices distributed over two reciprocal-space shells (Sec. S7 of the Supplemental Material [36]). Thus, the same pairwise sum- and difference-wave-vector mechanism that generates the two-scale $D_6$ superlattice extends naturally to the four-sublattice quasiperiodic order of the $D_8$ field.

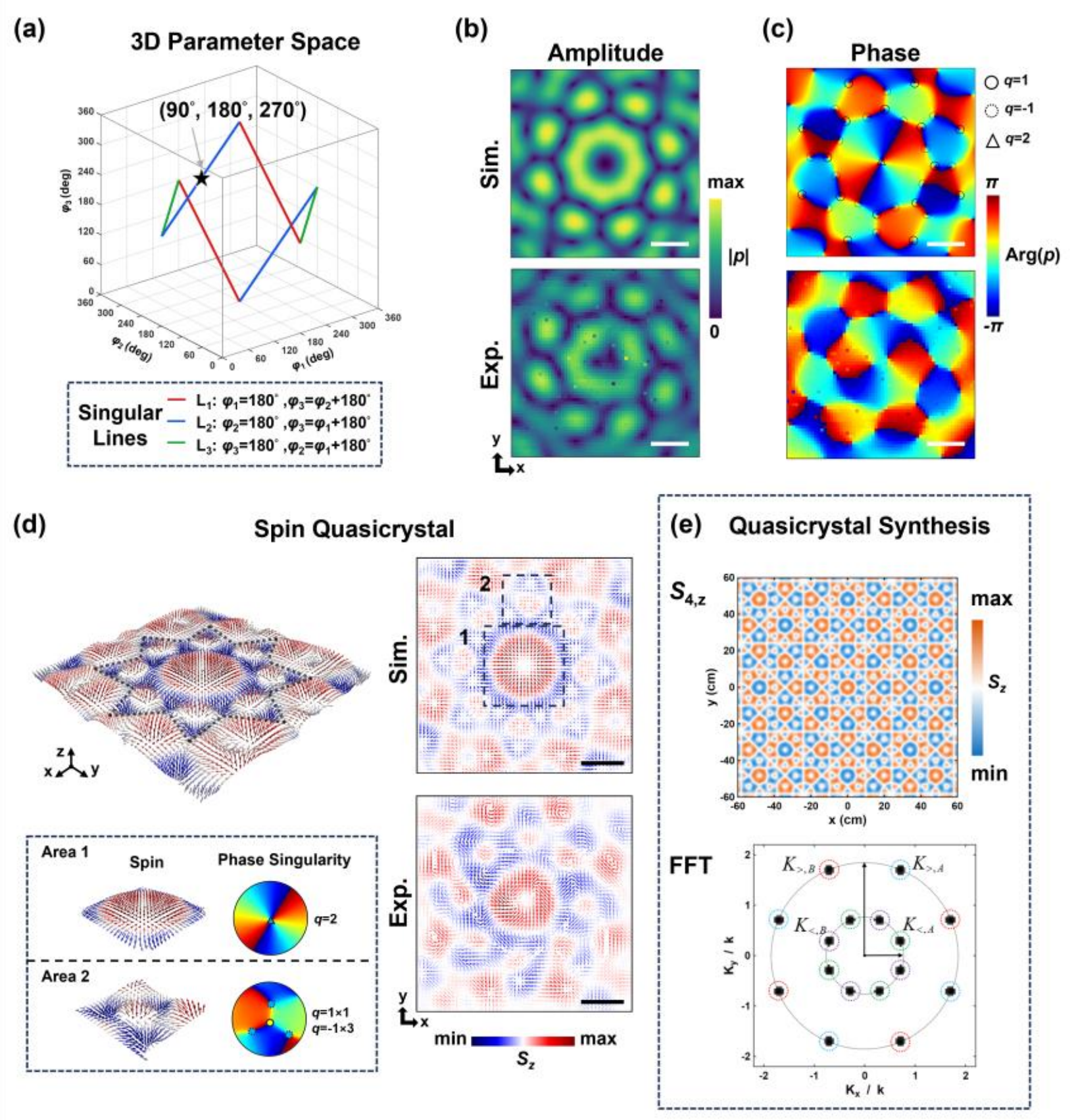

**FIG. 4. $D_8$ acoustic-spin quasicrystal.** (a) Three singular-line families of the central pressure in the ($\varphi_1$, $\varphi_2$, $\varphi_3$) parameter space; the star marks the selected encoding. (b), (c) Simulated (top) and measured (bottom) pressure amplitude and phase; symbols in (c) mark representative $q$=+1, $q$=-1, and $q$=+2 singularities (scale bars, 3 cm). (d) Simulated and reconstructed spin textures and representative local topological units; color denotes $S_z$, and arrows show the in-plane components (scale bars, 3 cm). (e) Analytical large-area $S_{4,z}$ and its FFT; colored markers group the reciprocal-space peaks into four $A$/$B$ subsets of the $K_<$ and $K_>$ shells.

***Bessel angular-spectrum limit and topology-composition map.***—The $D_4$ , $D_6$ , and $D_8$ configurations are discrete realizations of the polygonal standing-wave framework, although they need not share the same phase encoding. A finite $2N$ -sided field contains $N$ independent standing-wave channel orientations, which approach a continuous angular spectrum as $N$ increases. Starting from Eq. (7), with channel 0 chosen as the phase reference and equal normalized amplitudes, we impose the angular relative-phase encoding $\varphi_j = m\Theta_j$ . After removing an irrelevant unit-modulus global factor, the normalized discrete field can be written as

$$p_N^{(m)} = 2Ae^{-\tau z}\sum_{j=0}^{N-1} e^{im\Theta_j}\cos\left(k\hat{\mathbf{n}}_j\cdot\mathbf{r}_\parallel\right). \tag{23}$$

For $m = 0$ , all standing-wave channels are in phase and, in the continuous limit $N \to \infty$ ,

$$p_\infty^{(0)}(r) = J_0(kr). \tag{24}$$

Thus, the polygonal standing-wave field constitutes a normalized angular discretization of the zeroth-order Bessel spectrum. The corresponding finite $N$ field evolution and quantitative convergence are provided in Sec. S8 of the Supplemental Material [36].

Because the standing-wave basis is invariant under $\theta \to \theta + \pi$ , the angular weighting must be $\pi$ periodic. The lowest nontrivial angular encoding is therefore $m = 2$ . Its continuous limit is

$$p_\infty^{(2)}(r,\alpha) \propto -J_2(kr)e^{i2\alpha}. \tag{25}$$

where $\alpha$ is the polar angle. Near the origin, $J_2(kr) \propto r^2$ , so that the local field varies as $r^2e^{i2\alpha}$ , producing a $4\pi$ phase winding and a second-order singularity with $q = +2$ . The finite $D_6$ and $D_8$ fields reproduce this local Bessel-vortex core while remaining globally distinct: the former organizes its discrete interference harmonics into a double-skyrmion superlattice, whereas the latter superposes four meron/anti-meron sublattices into an eightfold acoustic-spin quasicrystal. The finite- $N$ correspondence and convergence toward the second-order Bessel limit are detailed in Sec. S9 of the Supplemental Material [36].

The resulting organization is summarized by the topology-composition map in Fig. 5, parameterized by the number $N$ of standing-wave channel orientations and the number $M_{\text{topo}}$ of distinguishable topological sublattices. The $D_4$ field forms a single meron/anti-meron lattice, the $D_6$ field contains two interpenetrating skyrmion sublattices, and the $D_8$ field comprises four meron/anti-meron sublattices with quasiperiodic order. For a fixed angular encoding, increasing $N$ continuously refines the angular sampling toward the corresponding Bessel spectrum. The discrete polygonal states and continuous Bessel fields are therefore unified by angular discretization, while their finite- $N$ long-range topology is determined by the symmetry and phase encoding of the available interference channels.

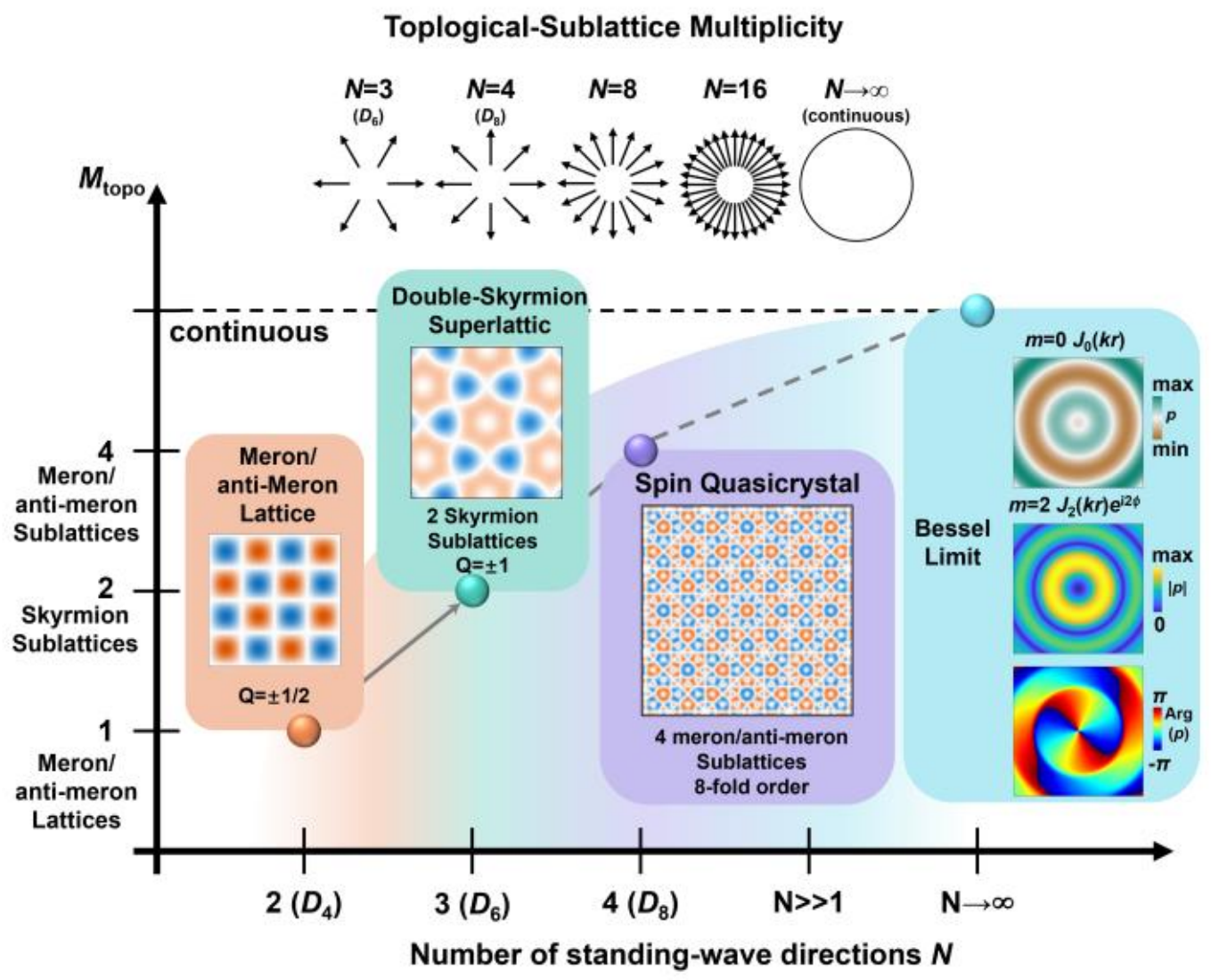

**FIG. 5. Topology-composition map of polygonal acoustic-spin textures.** The number of distinguishable topological sublattices $M_{\text{topo}}$ is shown versus the number $N$ of independent standing-wave channel orientations, connecting the $D_4$ meron/anti-meron lattice, $D_6$ double-skyrmion superlattice, and $D_8$ acoustic-spin quasicrystal with the continuous Bessel angular-spectrum limits.

***Conclusion.***—We have established a phase-singularity-mediated framework for constructing polygonal topological acoustic-spin states. Encoding the complex-pressure phase creates singularities that drive chiral time-averaged energy flux and, under SSAW confinement, time-stationary spin textures. The $D_4$ experiment verifies the mapping from alternating phase singularities to a meron/anti-meron lattice. In the $D_6$ configuration, parameter-space zeros select a $q = +2$ high-order phase core, and two spin subtextures of different scales form a double-skyrmion superlattice. In the $D_8$ configuration, the absence of a simple two-dimensional translational unit cell together with eightfold reciprocal-space peaks supports an acoustic spin quasicrystal composed of four meron/anti-meron sublattices. Finite polygonal fields can further be understood as discrete samples of a continuous Bessel angular spectrum: the $D_6$ and $D_8$ fields share a local $q = +2$ core but develop different long-range order because their discrete interference channels differ. This symmetry-phase-encoding-angular-discretization principle provides a general route to programmable topological acoustic-spin states.

Supplemental Material for

# "Phase-Singularity Control of Topological Acoustic Spin Textures"

Huaijin Ma[1†], Te Liu[1†], Jiachen Sheng[1], Xiaochang Pan[1], Wenwei Qian[1],
Xiangyu Chen[1], Kaiyuan Cao[1*], Jinpeng Yang[1*], Jian Wang[1*]

*1. College of Physical Science and Technology, Yangzhou University, Yangzhou 225002, China*

† These authors contributed equally to this work.
* Corresponding author: Kaiyuan Cao (kycao@yzu.edu.cn),
Jinpeng Yang (yangjp@yzu.edu.cn),
Jian Wang (wangjian@yzu.edu.cn).

## S1. Derivation of the time-averaged energy flux

We derive the time-averaged energy flux from the linearized Euler equation and show that it is governed by the phase gradient of the complex pressure. In a stationary, inviscid, lossless fluid, the particle velocity $\mathbf{u}(\mathbf{r})$ and pressure $p(\mathbf{r})$, with harmonic time dependence $e^{-i\omega t}$, satisfy the Euler equation. For angular frequency ω, the complex velocity is

$$\mathbf{u}(\mathbf{r}) = -\frac{i}{\omega\rho_0}\nabla p(\mathbf{r}), \tag{S1}$$

where $\rho_0$ is the mass density of the background medium and $i$ is the imaginary unit.

The time-averaged energy flux is

$$\mathbf{J}(\mathbf{r}) = \frac{1}{2}\mathrm{Re}\left[p^*(\mathbf{r})\mathbf{u}(\mathbf{r})\right]. \tag{S2}$$

where the asterisk denotes complex conjugation.

Writing the complex pressure in amplitude-phase form,

$$p(\mathbf{r}) = P_0(\mathbf{r})e^{i\phi(\mathbf{r})}, \tag{S3}$$

where $P_0(\mathbf{r})$ and $\phi(\mathbf{r})$ are the pressure amplitude and phase, respectively, the phase is

$$\phi(\mathbf{r}) = \mathrm{atan2}\left(\mathrm{Im}\left[p(\mathbf{r})\right],\mathrm{Re}\left[p(\mathbf{r})\right]\right). \tag{S4}$$

Taking the gradient of Eq. (S3), gives

$$\nabla p(\mathbf{r}) = e^{i\phi(\mathbf{r})}\left[\nabla P_0(\mathbf{r}) + iP_0(\mathbf{r})\nabla\phi(\mathbf{r})\right]. \tag{S5}$$

Substitution of Eq. (S5) into Eq. (S1), and then Eq. (S2), yields

$$\mathbf{J}(\mathbf{r}) = \frac{1}{2\omega\rho_0}P_0^2(\mathbf{r})\nabla\phi(\mathbf{r}), \tag{S6}$$

Equation (S6) shows that the local time-averaged energy flux follows the phase gradient of the complex pressure and is weighted by the squared local amplitude. A nonuniform phase therefore drives energy transport, while isolated pressure zeros organize the associated circulation topology.

## S2. Polygonal standing-wave theory

We generalize the orthogonal model to a polygonal excitation. The $2N$ boundary-launched SSAW components form $N$ counterpropagating pairs, each defining one standing-wave channel. Because an ideal standing-wave channel has no net propagation direction, only $N$ independent channel-axis orientations are required.

The in-plane rotation matrix is

$$R(\Theta)=\begin{pmatrix}\cos\Theta & -\sin\Theta \\ \sin\Theta & \cos\Theta\end{pmatrix}, \tag{S7}$$

where $\Theta$ is the rotation angle. The axis orientation of standing-wave channel $j$ is represented by

$$\hat{\mathbf{n}}_j = R(\Theta_j)\hat{\mathbf{e}}_x = \begin{pmatrix}\cos\Theta_j \\ \sin\Theta_j\end{pmatrix}, \tag{S8}$$

where $\hat{\mathbf{e}}_x = (1,0)$ is the x-directed unit vector and

$$\Theta_j = \frac{j\pi}{N}, j = 0,1...,N-1. \tag{S9}$$

Here $\hat{\mathbf{n}}_j$ is a representative unit vector along the channel axis, and $k$ denotes the in-plane SSAW wave number.

Each standing-wave channel is constructed from two equal-amplitude SSAW components propagating oppositely along the metasurface. We impose the same harmonic phase $\varphi_j$ on the two constituents of channel $j$ ,

$$p_j^{\pm}(\mathbf{r}_{\parallel},z) = A_j e^{i\varphi_j} e^{\pm ik\hat{\mathbf{n}}_j\cdot\mathbf{r}_{\parallel}} e^{-\tau z}. \tag{S10}$$

Their coherent superposition is

$$p_j(\mathbf{r}_{\parallel},z) = 2A_j e^{i\varphi_j}\cos\left[k\hat{\mathbf{n}}_j\cdot\mathbf{r}_{\parallel}\right]e^{-\tau z}. \tag{S11}$$

Thus, $\varphi_j$ is the common harmonic phase of the counterpropagating pair, rather than a propagation phase of the resulting standing wave. Because the two constituents share the same phase, the standing-wave spatial factor contains no additional phase offset. A phase mismatch within one counterpropagating pair would instead translate its nodal pattern; such an intrachannel phase degree of freedom is not considered here.

For compact notation, we define

$$c_j = A_j e^{i\varphi_j}, \tag{S12}$$

where $A_j$ is the constituent-wave amplitude and $\varphi_j$ is the common harmonic phase defined above. For an SSAW confined above the metasurface, the total complex pressure generated by the $N$ channels is

$$p_N(\mathbf{r}_{\parallel},z) = 2e^{-\tau z}\sum_{j=0}^{N-1} c_j\cos u_j, \tag{S13}$$

where $\mathbf{r}_{\parallel} = (x,y)$ , $u_j = k\hat{\mathbf{n}}_j\cdot\mathbf{r}_{\parallel} = k(x\cos\Theta_j + y\sin\Theta_j)$ , and $\tau = \sqrt{k^2-k_0^2} > 0$ is the evanescent decay constant normal to the surface; $k_0 = \omega/c_0$ is the free-space acoustic wave number. A uniform phase shift $\varphi_j \to \varphi_j + \varphi_g$ applied to all channels multiplies $p_N$ by the common factor $e^{i\varphi_g}$ and therefore leaves the pressure amplitude, phase-singularity topology, time-averaged energy flux, and acoustic spin unchanged. We may therefore choose channel 0 as the phase reference $\varphi_0 = 0$ . An $N$ -channel field then contains $N-1$ independent relative phase degrees of freedom.

The pressure in Eq. (S13) can be decomposed as

$$p_N(\mathbf{r}_{\parallel},z) = R_N(\mathbf{r}_{\parallel},z) + iI_N(\mathbf{r}_{\parallel},z), \tag{S14}$$

with $R_N(\mathbf{r}_{\parallel},z) = 2e^{-\tau z}\sum_{j=0}^{N-1} A_j\cos\varphi_j\cos u_j$ and $I_N(\mathbf{r}_{\parallel},z) = 2e^{-\tau z}\sum_{j=0}^{N-1} A_j\sin\varphi_j\cos u_j$ . The local pressure phase is

$$\phi_N(\mathbf{r}_\parallel)=\operatorname{atan2}\left[I_N(\mathbf{r}_\parallel,z),R_N(\mathbf{r}_\parallel,z)\right]. \tag{S15}$$

We characterize the phase structure using the effective phase-gradient field

$$\mathbf{A}_{eff}=\nabla_\parallel\phi_N. \tag{S16}$$

where $\nabla_\parallel=(\partial x,\partial y)$. Away from pressure zeros, it can be evaluated without explicit phase unwrapping as

$$\mathbf{A}_{eff}=\frac{R_N\nabla_\parallel I_N-I_N\nabla_\parallel R_N}{R_N^2+I_N^2}. \tag{S17}$$

At a pressure zero, $R_N=I_N=0$, both the phase and $\mathbf{A}_{eff}$ are singular. The charge enclosed by a closed contour $C$ is

$$q=\frac{1}{2\pi}\oint_C\mathbf{A}_{eff}\cdot d\mathbf{l}. \tag{S18}$$

Unless stated otherwise, $C$ is counterclockwise when viewed from +z.

**A. Time-averaged energy flux**

We derive the time-averaged energy flux from Eq. (S2):

$$\mathbf{J}_N=\frac{1}{2}\operatorname{Re}\left[p_N^*\mathbf{u}_N\right]. \tag{S19}$$

The complex particle velocity satisfies

$$\mathbf{u}_N=-\frac{i}{\omega\rho_0}\nabla p_N. \tag{S20}$$

Using $p_N=R_N+iI_N$, Eq. (S19) becomes

$$\mathbf{J}_N=\frac{1}{2\rho_0\omega}\left(R_N\nabla I_N-I_N\nabla R_N\right). \tag{S21}$$

The evanescent factor $e^{-\tau z}$ contributes no vertical phase gradient. Considering the effect of $\nabla_\parallel=(\partial x,\partial y)$, we find

$$\begin{aligned}R_N\nabla I_N&=-4e^{-2\tau z}k\sum_{j=0}^{N-1}\sum_{l=0}^{N-1}A_jA_l\cos\varphi_j\sin\varphi_l\cos u_j\sin u_l\hat{\mathbf{n}}_l,\\ I_N\nabla_\parallel R_N&=-4e^{-2\tau z}k\sum_{j=0}^{N-1}\sum_{l=0}^{N-1}A_jA_l\sin\varphi_j\cos\varphi_l\cos u_j\sin u_l\hat{\mathbf{n}}_l.\end{aligned} \tag{S22}$$

Substitution of Eq. (S22) into Eq. (S21) yields

$$\mathbf{J}_{N,\parallel}=-\frac{2e^{-2\tau z}k}{\rho_0\omega}\sum_{j=0}^{N-1}\sum_{l=0}^{N-1}A_jA_l\sin(\varphi_l-\varphi_j)\cos u_j\sin u_l\hat{\mathbf{n}}_l. \tag{S23}$$

To separate the angular-channel contributions, we group the indices $(j,l)$ and $(l,j)$ and define $\Delta\varphi_{jl}=\varphi_l-\varphi_j$. The double sum contains the cases $j=l$, $j<l$, and $j>l$.

For $j=l$, $\Delta\varphi_{jj}=\varphi_j-\varphi_j=0$, so all self-terms vanish: $\mathbf{J}_{j,j}=0$. A single standing-wave channel therefore carries no net active flux; nonzero flux arises from phase-sensitive interference between distinct channels.

For distinct channels $j\neq l$, the double sum contains both the $(j,l)$ and $(l,j)$ terms. The $(j,l)$ term is

$$\mathbf{J}_{j,l}=-\frac{2e^{-2\tau z}k}{\rho_0\omega}A_jA_l\sin(\varphi_l-\varphi_j)\cos u_j\sin u_l\hat{\mathbf{n}}_l. \tag{S24}$$

whereas exchanging the indices gives

$$\mathbf{J}_{l,j}=-\frac{2e^{-2\tau z}k}{\rho_0\omega}A_lA_j\sin(\varphi_j-\varphi_l)\cos u_l\sin u_j\hat{\mathbf{n}}_j. \tag{S25}$$

Using $A_jA_l=A_lA_j$ and $\sin(\varphi_j-\varphi_l)=-\sin(\varphi_l-\varphi_j)=-\sin\Delta\varphi_{jl}$, Eqs. (S24) and (S25) become

$$\mathbf{J}_{j,l} = -\frac{2e^{-2\tau z}k}{\rho_0\omega} A_j A_l \sin\Delta\varphi_{jl} \cos u_j \sin u_l \hat{\mathbf{n}}_l,$$
$$\mathbf{J}_{l,j} = \frac{2e^{-2\tau z}k}{\rho_0\omega} A_j A_l \sin\Delta\varphi_{jl} \cos u_l \sin u_j \hat{\mathbf{n}}_j. \tag{S26}$$

Adding the exchanged terms gives

$$\mathbf{J}_{j,l} + \mathbf{J}_{l,j} = \frac{2e^{-2\tau z}k}{\rho_0\omega} A_j A_l \sin\Delta\varphi_{jl}\left(\cos u_l \sin u_j \hat{\mathbf{n}}_j - \cos u_j \sin u_l \hat{\mathbf{n}}_l\right). \tag{S27}$$

The original double sum decomposes as

$$\sum_{j=0}^{N-1}\sum_{l=0}^{N-1} = \sum_{j=l} + \sum_{j<l} + \sum_{j>l}, \tag{S28}$$

where the self-terms $j = l$ vanish, and the $j > l$ terms combine with their exchanged counterparts in the $j < l$ sector. Thus, the total in-plane flux consists entirely of pairwise-interference channels:

$$\mathbf{J}_{N,\parallel} = \frac{2e^{-2\tau z}k}{\rho_0\omega} \sum_{j<l} A_j A_l \sin\Delta\varphi_{jl}\left(\cos u_l \sin u_j \hat{\mathbf{n}}_j - \cos u_j \sin u_l \hat{\mathbf{n}}_l\right). \tag{S29}$$

Equation (S29) confirms that distinct channel pairs generate the in-plane energy flux. The interchannel phase factor $\sin\Delta\varphi_{jl}$ controls the sign and magnitude of each pairwise contribution. For $\Delta\varphi_{jl} = \ell\pi, \ell \in \mathbb{Z}$, that pair contributes no active flux; for $\Delta\varphi_{jl} \neq \ell\pi$, it produces a directional interference flux. Reversing $\Delta\varphi_{jl}$ reverses that channel's contribution, whose magnitude is maximal for $\left|\sin\Delta\varphi_{jl}\right| = 1$. This explains the $D_4$ choice $\Delta\varphi_{10} = \pi/2$ and, more generally, the control of flux chirality and spin polarity by interchannel phase differences. Reversing the complete flux pattern requires complex conjugation of the full phase encoding, not reversal of a single pairwise phase difference.

**B. Spatial-harmonic decomposition**

To expose the spatial harmonics of the flux, we use

$$\sin u_j \cos u_l = \frac{1}{2}\left(\sin\left(u_j + u_l\right) + \sin\left(u_j - u_l\right)\right),$$
$$\sin u_l \cos u_j = \frac{1}{2}\left(\sin\left(u_j + u_l\right) - \sin\left(u_j - u_l\right)\right). \tag{S30}$$

Substitution of Eq. (S30) into Eq. (S29) gives

$$\mathbf{J}_{N,\parallel} = \frac{e^{-2\tau z}k}{\rho_0\omega} \sum_{j<l} A_j A_l \sin\Delta\varphi_{jl}\left(\left(\hat{\mathbf{n}}_j - \hat{\mathbf{n}}_l\right)\sin\left(u_j + u_l\right) + \left(\hat{\mathbf{n}}_j + \hat{\mathbf{n}}_l\right)\sin\left(u_j - u_l\right)\right). \tag{S31}$$

Because $u_j \pm u_l = k\left(\hat{\mathbf{n}}_j \pm \hat{\mathbf{n}}_l\right)\cdot \mathbf{r}_\parallel$, we define the sum and difference wave vectors

$$\mathbf{K}_{jl}^{+} = k\left(\hat{\mathbf{n}}_j + \hat{\mathbf{n}}_l\right),$$
$$\mathbf{K}_{jl}^{-} = k\left(\hat{\mathbf{n}}_j - \hat{\mathbf{n}}_l\right). \tag{S32}$$

The flux then takes the spatial-harmonic form

$$\mathbf{J}_{N,\parallel} = \frac{e^{-2\tau z}k}{\rho_0\omega} \sum_{j<l} A_j A_l \sin\Delta\varphi_{jl}\left(\left(\hat{\mathbf{n}}_j - \hat{\mathbf{n}}_l\right)\sin\left(\mathbf{K}_{jl}^{+}\cdot \mathbf{r}\right) + \left(\hat{\mathbf{n}}_j + \hat{\mathbf{n}}_l\right)\sin\left(\mathbf{K}_{jl}^{-}\cdot \mathbf{r}\right)\right). \tag{S33}$$

Thus, every pair of standing-wave channel orientations contributes modulation through both the sum ($\mathbf{K}_{jl}^{+}$) and difference ($\mathbf{K}_{jl}^{-}$) wave vectors.

From Eq. (S9), the angular separation between two channel orientations is

$$\Delta\Theta_{jl} = \Theta_l - \Theta_j = \frac{\left(l - j\right)\pi}{N}. \tag{S34}$$

For $j < l$, we define the angular-interference-channel index

$$s = l - j, \tag{S35}$$

so that

$$\Delta\Theta_s = \frac{s\pi}{N}. \tag{S36}$$

The geometry of the two unit orientation vectors gives

$$\begin{aligned} \hat{\mathbf{n}}_j + \hat{\mathbf{n}}_l &= 2\cos\left(\frac{\Delta\Theta_s}{2}\right) R\left(\frac{\Theta_j + \Theta_l}{2}\right)\hat{\mathbf{e}}_x, \\ \hat{\mathbf{n}}_j - \hat{\mathbf{n}}_l &= 2\sin\left(\frac{\Delta\Theta_s}{2}\right) R\left(\frac{\Theta_j + \Theta_l}{2} - \frac{\pi}{2}\right)\hat{\mathbf{e}}_x, \end{aligned} \tag{S37}$$

Hence, the interference-wave-vector magnitudes are

$$\begin{aligned} \left|\mathbf{K}_{jl}^{+}\right| &= 2k\cos\left(\frac{s\pi}{2N}\right), \\ \left|\mathbf{K}_{jl}^{-}\right| &= 2k\sin\left(\frac{s\pi}{2N}\right). \end{aligned} \tag{S38}$$

Here $s = 1, 2, ..., N-1$ labels the angular-interference channel. Increasing $N$ introduces additional channel orientations and spatial harmonics, allowing the flux and spin to evolve from simple alternating patterns to multichannel, multiscale, multinodal topological textures.

**C. Acoustic-spin field**

In a homogeneous, source-free, inviscid region, the spin density is related to the time-averaged energy flux by

$$\mathbf{S}_N = \frac{1}{\omega^2}\nabla \times \mathbf{J}_N. \tag{S39}$$

For one pairwise channel in Eq. (S29), write

$$\begin{aligned} \mathbf{J}_{jl,\parallel} &= C_{jl}\left(\cos u_l \sin u_j \hat{\mathbf{n}}_j - \cos u_j \sin u_l \hat{\mathbf{n}}_l\right), \\ C_{jl} &= \frac{2e^{-2\tau z}k}{\rho_0\omega} A_j A_l \sin\Delta\varphi_{jl}. \end{aligned} \tag{S40}$$

Retaining the in-plane derivatives isolates the out-of-plane component associated with in-plane circulation:

$$S_{jl,z} = \frac{C_{jl}}{\omega^2}\left(\left(\nabla_\parallel \cos u_l \sin u_j\right)\times\hat{\mathbf{n}}_j - \left(\nabla_\parallel \cos u_j \sin u_l\right)\times\hat{\mathbf{n}}_l\right)_z. \tag{S41}$$

Using $u_j = k\hat{\mathbf{n}}_j \cdot \mathbf{r}, u_l = k\hat{\mathbf{n}}_l \cdot \mathbf{r}$, Eq. (S41) becomes

$$S_{jl,z} = \frac{4e^{-2\tau z}k^2}{\rho_0\omega^3} A_j A_l \sin\Delta\varphi_{jl}\left(\hat{\mathbf{n}}_j \times \hat{\mathbf{n}}_l\right)_z \sin u_l \sin u_j. \tag{S42}$$

The out-of-plane spin therefore originates from the nonparallel geometry of the interfering standing-wave channels. Using $\left(\hat{\mathbf{n}}_j \times \hat{\mathbf{n}}_l\right)_z = \sin\left(\Theta_l - \Theta_j\right)$, we obtain

$$S_{jl,z} = \frac{4e^{-2\tau z}k^2}{\rho_0\omega^3} A_j A_l \sin\Delta\varphi_{jl} \sin\left(\Theta_l - \Theta_j\right)\sin u_l \sin u_j. \tag{S43}$$

Omitting the common positive amplitude prefactor, its spatial and phase dependence is

$$S_{jl,z} \propto \sin\Delta\varphi_{jl} \sin\left(\Theta_l - \Theta_j\right)\sin u_l \sin u_j. \tag{S44}$$

A pairwise angular channel is therefore governed by three factors:

$$S_{jl,z} \sim \sin\left[\text{Phase difference}\right]\times\sin\left[\text{Geometric angle}\right]\times\left[\text{Spatial harmonics}\right]. \tag{S45}$$

The phase factor $\sin\Delta\varphi_{jl}$ is necessary for chiral flux and nonzero out-of-plane spin. For $\Delta\varphi_{jl} = \ell\pi, \ell \in \mathbb{Z}$, the pair contributes nothing to $S_{jl,z}$. Reversing $\Delta\varphi_{jl}$ reverses this channel's spin contribution, whereas reversing the entire texture requires conjugating the full encoding. The geometric factor $\sin(\Theta_l - \Theta_j)$ quantifies the relative orientation of the two nonparallel channel axes. Under the convention $j < l$, $0 < \Theta_l - \Theta_j < \pi$, so it is positive and its magnitude sets the coupling strength. The spatial factor $\sin u_l \sin u_j$ fixes the sign distribution and nodes of the spin channel. The product of the two standing-wave functions determines the local modulation of $S_{jl,z}$. This differs from the flux magnitude itself because spin measures the local rotational character of the energy flow.

Thus, a nonzero pairwise out-of-plane spin contribution requires a nontrivial interchannel phase difference, nonparallel channel orientations, and nonvanishing local standing-wave modulation. Summing all pairs gives the total out-of-plane spin:

$$S_{N,z} = \frac{4e^{-2\tau z}k^2}{\rho_0 \omega^3} \sum_{j<l} A_j A_l \sin\Delta\varphi_{jl} \sin(\Theta_l - \Theta_j) \sin u_l \sin u_j. \tag{S46}$$

Equation (S46) shows that increasing N expands the set of angular-interference channels. Their spatial harmonics and geometric couplings transform simple alternating lattices into multichannel, multiscale, multinodal textures, providing a unified description of fourfold, sixfold, eightfold, and higher polygonal fields.

For illustration, $N = 2$, $N = 3$ and $N = 4$ correspond to $D_4$, $D_6$ and $D_8$ standing-wave constraints. The corresponding channel-axis orientations are

$$D_4 : \hat{\mathbf{n}}_0 = (1,0), \hat{\mathbf{n}}_0 = (0,1). \tag{S47}$$

$$D_6 : \hat{\mathbf{n}}_0 = (1,0), \hat{\mathbf{n}}_1 = \left(\frac{1}{2}, \frac{\sqrt{3}}{2}\right), \hat{\mathbf{n}}_2 = \left(-\frac{1}{2}, \frac{\sqrt{3}}{2}\right). \tag{S48}$$

and

$$D_8 : \hat{\mathbf{n}}_0 = (1,0), \hat{\mathbf{n}}_1 = \left(\frac{1}{\sqrt{2}}, \frac{1}{\sqrt{2}}\right), \hat{\mathbf{n}}_2 = (0,1), \hat{\mathbf{n}}_3 = \left(-\frac{1}{\sqrt{2}}, \frac{1}{\sqrt{2}}\right). \tag{S49}$$

For equal amplitudes $A_j = A$ and with channel 0 chosen as the phase reference $\varphi_0 = 0$, Eq. (S13) reduces to the following forms.

For the $D_4$ configuration,

$$\begin{aligned} p_2(x,y,z) &= 2Ae^{-\tau z} \sum_{j=0}^{1} e^{i\varphi_j} \cos\left[k\left(x\cos\Theta_j + y\sin\Theta_j\right)\right], \\ p_2(x,y,z) &= 2Ae^{-\tau z}\left(\cos(kx) + e^{i\varphi_1}\cos(ky)\right). \end{aligned} \tag{S50}$$

For the $D_6$ configuration,

$$\begin{aligned} p_3(x,y,z) &= 2Ae^{-\tau z} \sum_{j=0}^{2} e^{i\varphi_j} \cos\left[k\left(x\cos\Theta_j + y\sin\Theta_j\right)\right], \\ p_3(x,y,z) &= 2Ae^{-\tau z}\left(\cos(kx) + e^{i\varphi_1}\cos\left[\frac{1}{2}k\left(x+\sqrt{3}y\right)\right] + e^{i\varphi_2}\cos\left[\frac{1}{2}k\left(x-\sqrt{3}y\right)\right]\right). \end{aligned} \tag{S51}$$

For the $D_8$ configuration,

$$\begin{aligned} p_4(x,y,z) &= 2Ae^{-\tau z} \sum_{j=0}^{3} e^{i\varphi_j} \cos\left[k\left(x\cos\Theta_j + y\sin\Theta_j\right)\right], \\ p_4(x,y,z) &= 2Ae^{-\tau z}\left(\cos(kx) + e^{i\varphi_1}\cos\left[\frac{k(x+y)}{\sqrt{2}}\right] + e^{i\varphi_2}\cos(ky) + e^{i\varphi_3}\cos\left[\frac{k(-x+y)}{\sqrt{2}}\right]\right). \end{aligned} \tag{S52}$$

The Time-average energy flux and acoustic spin under the three types of symmetry are:

$$\mathbf{J}_{2,\|} = \frac{2e^{-2\tau z}k}{\rho_0\omega}\sum_{j<l} A_j A_l \sin\Delta\varphi_{jl}\left(\cos u_l \sin u_j \hat{\mathbf{n}}_j - \cos u_j \sin u_l \hat{\mathbf{n}}_l\right),$$
$$\mathbf{J}_{3,\|} = \frac{2e^{-2\tau z}k}{\rho_0\omega}\sum_{j<l} A_j A_l \sin\Delta\varphi_{jl}\left(\cos u_l \sin u_j \hat{\mathbf{n}}_j - \cos u_j \sin u_l \hat{\mathbf{n}}_l\right), \tag{S53}$$
$$\mathbf{J}_{4,\|} = \frac{2e^{-2\tau z}k}{\rho_0\omega}\sum_{j<l} A_j A_l \sin\Delta\varphi_{jl}\left(\cos u_l \sin u_j \hat{\mathbf{n}}_j - \cos u_j \sin u_l \hat{\mathbf{n}}_l\right).$$

and

$$S_{2,z} = \frac{4e^{-2\tau z}k^2}{\rho_0\omega^3}\sum_{j<l} A_j A_l \sin\Delta\varphi_{jl} \sin\left(\Theta_l - \Theta_j\right)\sin u_l \sin u_j,$$
$$S_{3,z} = \frac{4e^{-2\tau z}k^2}{\rho_0\omega^3}\sum_{j<l} A_j A_l \sin\Delta\varphi_{jl} \sin\left(\Theta_l - \Theta_j\right)\sin u_l \sin u_j, \tag{S54}$$
$$S_{4,z} = \frac{4e^{-2\tau z}k^2}{\rho_0\omega^3}\sum_{j<l} A_j A_l \sin\Delta\varphi_{jl} \sin\left(\Theta_l - \Theta_j\right)\sin u_l \sin u_j.$$

## S3. Experimental details and acoustic-spin reconstruction

The acoustic metasurfaces were modeled in COMSOL Multiphysics. The basic element is a locally resonant circular-hole unit cell with lattice constant a, hole radius r, and depth h [Fig. S1(a)]. The depth primarily sets the local resonance and surface-wave dispersion, whereas $r/a$ controls coupling between the cavities and surrounding air.

For the $D_4$ configuration, the metasurface is a 30×30 square array of identical cells [Fig. S1(b)] with $a$=10 mm, $r$=4 mm, and $h$=28 mm. For the $D_6$ and $D_8$ configurations, a disk-shaped metasurface of radius 20 cm accommodates multidirectional boundary excitation [Fig. S1(e)]. It uses the same circular-hole cells, with $a$=10 mm, $r$=3.3 mm, and $h$=24 mm.

The circular-hole metasurface supports an SSAW mode. Its in-plane wave number can be estimated from the effective-impedance dispersion relation

$$k_{ssaw} = k_0\sqrt{1+\pi^2\left(\frac{r}{a}\right)^4 \tan^2\left(k_0 h\right)}, \tag{S55}$$

where $k_0 = 2\pi f / c_0$ is the free-space wave number in air, $f$ is the operating frequency, and $c_0$ is the sound speed in air. Equation (S55) follows by matching the effective surface impedance of the closed-hole array to the impedance of an evanescent wave in air.

For the $D_4$ configuration at $f$=2040 Hz, substitution of the structural parameters into Eq. (S55) gives $k_{ssaw} = 49.51\ \mathrm{m}^{-1}$. Because $k_{ssaw} > k_0$, the field decays exponentially normal to the surface. We further calculated the surface-mode dispersion of the $D_4$ unit cell and compared it with the sound cone [Fig. S1(c)]. At 2040 Hz, the simulated branch gives $k_{ssaw} = 49.31\ \mathrm{m}^{-1}$, close to the analytical estimate from Eq. (S55). The corresponding eigenmode [Fig. S1(d)] decays rapidly with z above the holes, confirming evanescent surface confinement. Thus, 2040 Hz is a self-consistent operating frequency for exciting the SSAW in the $D_4$ experiment.

For the $D_6$ and $D_8$ configurations at $f$=2700 Hz, Eq. (S55) gives the analytical estimate $k_{ssaw} = 64.83\ \mathrm{m}^{-1}$. The simulated branch gives $k_{ssaw} = 66.35\ \mathrm{m}^{-1}$ at 2700 Hz [Fig. S1(f)], close to the analytical result, and the eigenmode remains evanescent [Fig. S1(g)]. Accordingly, the $D_4$ experiment uses f=2040 Hz, while the $D_6$ and $D_8$ experiments use $f$=2700 Hz.

The samples were fabricated by three-dimensional printing. Signals from a 33500B generator (Keysight) were amplified by a PX3 power amplifier (Yamaha) and drove four loudspeaker arrays whose phases and amplitudes matched the numerical design. The complex pressure was measured on two parallel planes above the metasurface using a 378C01 probe (PCB Piezotronics) mounted on a three-dimensional scanning stage with 5-mm lateral steps. The two planes provide the normal pressure gradient required to reconstruct the three-dimensional velocity of the SSAW near field.

The acoustic spin was reconstructed from the simulated or measured complex pressure. We write $p = Pe^{i\phi}$, where P and $\phi$ are the measured amplitude and phase. The velocity follows from $\mathbf{u} = \left(i\omega\rho_0\right)^{-1}\nabla p$. In simulation, all derivatives were evaluated directly. Experimentally, lateral pressure gradients gave the in-plane velocity components, and the pressure difference between the two planes approximated the normal derivative and hence the out-of-plane velocity. This procedure reconstructs the full three-dimensional complex velocity above the surface.

We then calculated $\mathbf{S} = \left(\rho_0 / 2\omega\right)\mathrm{Im}\left(\mathbf{u}^* \times \mathbf{u}\right)$ and used the normalized spin field $\mathbf{n} = \mathbf{S} / |\mathbf{S}|$ to evaluate the skyrmion density and number. Two-plane reconstruction is essential because both lateral and normal velocity components enter the three-dimensional spin texture.

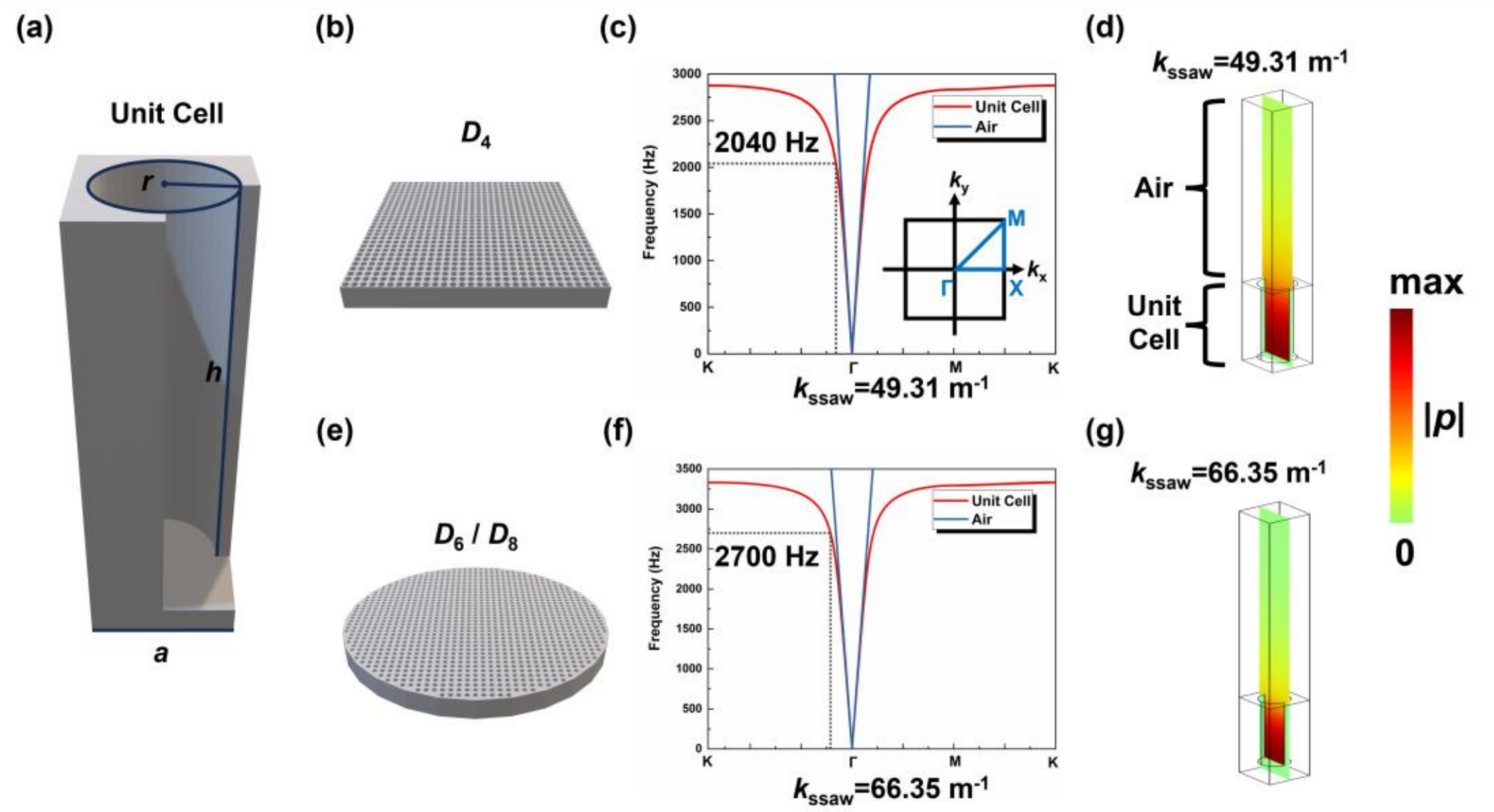

**FIG. S1. SSAW-mode verification for the $D_4$ and $D_6/D_8$ metasurfaces.** (a) Circular-hole resonant unit cell. (b) Square perforated metasurface used for $D_4$. (c) $D_4$ surface-mode dispersion; the dashed line marks $f$=2040 Hz and $k_{SSAW}$=49.31 m$^{-1}$. (d) Corresponding eigenmode, showing evanescent confinement above the unit cell. (e) Disk-shaped perforated metasurface used for $D_6/D_8$. (f) $D_6/D_8$ surface-mode dispersion; the dashed line marks $f$=2700 Hz and $k_{SSAW}$=66.35 m$^{-1}$. (g) Corresponding evanescent eigenmode.

## S4. Phase singularities, topological charges and chiral energy-flux for $D_4$-symmetry

Following the construction in Sec. S2, the $D_4$ field consists of two orthogonal standing-wave channels, each formed by an equal-phase counterpropagating pair. We choose the $x$ oriented channel as the phase reference $\varphi_0 = 0$, and denote the phase offset of the $y$ oriented channel by $\varphi_1$. Hence, $\varphi_1 = \Delta\varphi_{01}$ is the interchannel phase difference between the two orthogonal channels. With constituent amplitudes $A$ and $B$, the synthesized field is

$$p_2(x,y) = 2\left[A\cos(kx) + B\cos\varphi_1\cos(ky)\right] + i2B\sin\varphi_1\cos(ky), \tag{S56}$$

The synthesized field has real and imaginary parts $R = 2\left[A\cos(kx) + B\cos\varphi_1\cos(ky)\right]$ and $I = 2B\sin\varphi_1\cos(ky)$, respectively. Writing the complex pressure as $p(x,y) = P_2(x,y)e^{i\phi_2(x,y)}$, its amplitude and phase are

$$\begin{aligned} P_2(x,y) &= \sqrt{R^2(x,y) + +I^2(x,y)}, \\ \phi_2(x,y) &= \operatorname{atan2}\left[I(x,y), R(x,y)\right]. \end{aligned} \tag{S57}$$

We consider two representative limits. When $\varphi_1 = \ell\pi(\ell \in \mathbb{Z})$, $I = 0$, and the synthesized field is purely real. Wherever its amplitude is nonzero, the phase is piecewise constant at 0 or $\pi$ (mod $2\pi$), subject to the branch convention. It changes discontinuously only across the nodal lines $p_2(x,y) = 0$ and has zero gradient elsewhere. Such a real standing-wave field cannot sustain a nonzero time-averaged energy flux.

By contrast, for $\sin\varphi_1 \neq 0$, the pressure is genuinely complex and can possess isolated zeros with quantized phase winding. These singularities generate the phase gradients that drive the energy flux discussed in the main text. We next determine their positions and charges. A nontrivial complex phase enables phase singularities and phase gradients [1,2]. The phase $\phi_2(x,y)$ becomes undefined at pressure zeros [3]. For Eq. (S57), this requires the simultaneous conditions

$$R(x,y) = 0, \quad I(x,y) = 0. \tag{S58}$$

For nonzero $A$, $B$, and $\sin\varphi_1 \neq 0$, the second condition in Eq. (S58) gives

$$\cos(kx) = 0, \quad \cos(ky) = 0. \tag{S59}$$

Thus, the phase singularities are located at

$$(x_\mu, y_\nu) = \left(\frac{\frac{\pi}{2} + \mu\pi}{k}, \frac{\frac{\pi}{2} + \nu\pi}{k}\right), \quad \mu, \nu \in \mathbb{Z}. \tag{S60}$$

For $\sin\varphi_1 = 0$, by contrast, the isolated complex zeros merge into nodal lines of a real standing wave. The phase becomes piecewise constant, and quantized winding around the former isolated zeros is lost.

The phase topological charge $q$ can be defined for a closed loop $C$ enclosing only one isolated phase singularity and avoiding all others as

$$q = \frac{1}{2\pi}\oint_C \nabla\phi(x,y)\cdot d\mathbf{l}. \tag{S61}$$

The contour $C$ is oriented counterclockwise when viewed from $+z$. To evaluate the charge of the singularity at $(x_\mu, y_\nu)$, we expand the coordinates locally as

$$x = x_\mu + \delta x, \quad y = y_\nu + \delta y, \quad |\delta x|, |\delta y| \ll k^{-1}. \tag{S62}$$

The pressure near the singularity $p_s$ is then

$$p_s = 2\left[A\cos\left(\frac{\pi}{2} + \mu\pi + k\delta x\right) + e^{i\varphi_1}B\cos\left(\frac{\pi}{2} + \nu\pi + k\delta y\right)\right]. \tag{S63}$$

Simplified to

$$p_s \approx -2k\left[A(-1)^{\mu}\delta x + e^{i\varphi_1}B(-1)^{\nu}\delta y\right]. \tag{S64}$$

On a small circular loop around the singularity,

$$\delta x = r\cos\alpha,\qquad \delta y = r\sin\alpha. \tag{S65}$$

we obtain

$$p_s(\alpha) \approx -2kr\left[A(-1)^{\mu}\cos\alpha + e^{i\varphi_1}B(-1)^{\nu}\sin\alpha\right], \tag{S66}$$

Introducing $C_0 = -2kr(-1)^{\mu}$ and $C = B(-1)^{\mu+\nu}$ simplifies Eq. (S66) to

$$p_s(\alpha) \approx C_0\left[A\cos\alpha + C\sin\alpha\cos\varphi_1 + iC\sin\alpha\sin\varphi_1\right], \tag{S67}$$

Writing its real and imaginary parts as $R_S = C_0(A\cos\alpha + C\sin\alpha\cos\varphi_1)$ and $I_S = C_0 C\sin\alpha\sin\varphi_1$, respectively, the local phase gradient follows from Eq. (S57):

$$\begin{aligned}\nabla\phi(\alpha) &= \frac{\partial}{\partial\alpha}\operatorname{atan2}\left(I(\alpha),R(\alpha)\right),\\ \nabla\phi(\alpha) &= \frac{AC\sin\varphi_1}{A^2\cos^2\alpha + C^2\sin^2\alpha + 2AC\sin\alpha\cos\alpha\cos\varphi_1}.\end{aligned} \tag{S68}$$

Substitution into Eq. (S61),, together with the π periodicity of Eq. (S68) , gives

$$q = \frac{1}{\pi}\int_0^{\pi}\frac{AC\sin\varphi_1}{A^2\cos^2\alpha + C^2\sin^2\alpha + 2AC\sin\alpha\cos\alpha\cos\varphi_1}d\alpha, \tag{S69}$$

Defining $t = \tan\alpha$, we obtain

$$\begin{aligned}q &= \frac{1}{\pi}AC\sin\varphi_1\int_{-\infty}^{\infty}\frac{dt}{C^2\left(t+\frac{A}{C}\cos\theta\right)^2 + A^2\sin^2\theta},\\ q &= \frac{1}{\pi}AC\sin\varphi_1\frac{\pi}{\left|AC\sin\varphi_1\right|},\\ q_{m,n} &= \operatorname{sgn}\left(AB\sin\varphi_1\right)(-1)^{\mu+\nu},\qquad \mu,\nu\in\mathbb{Z}.\end{aligned} \tag{S70}$$

Equation (S70) shows that a finite interchannel phase difference between the two orthogonal standing-wave channels produces a regular lattice of $q$=±1 singularities. Their alternating signs form the phase-singularity lattice that underlies the chiral energy-flux lattice and acoustic-spin meron texture discussed in the main text.

According to Eq. (S6), the time-averaged energy flux $\mathbf{J}(\mathbf{r})$ depends on the phase gradient $\nabla\phi$ and squared amplitude $P^2$. Substituting Eq. (S57) gives the two-dimensional flux

$$\begin{aligned}\mathbf{J}(x,y) &= \frac{1}{2\omega\rho_0}\left(R(x,y)\nabla I(x,y) - I(x,y)\nabla R(x,y)\right),\\ \mathbf{J}(x,y) &= \frac{2ABk\sin\varphi_1}{\omega\rho_0}\left(\sin(kx)\cos(ky), -\cos(kx)\sin(ky)\right).\end{aligned} \tag{S71}$$

Because this two-dimensional flux is divergence-free, $\nabla\cdot\mathbf{J} = 0$, a stream function $\psi(x,y)$ [4] can represent its circulation topology through

$$\begin{cases} J_x = \dfrac{\partial}{\partial y}\psi(x,y),\\ J_y = -\dfrac{\partial}{\partial x}\psi(x,y).\end{cases} \tag{S72}$$

where

$$\psi(x,y) = \frac{2AB\sin\varphi_1}{\omega\rho_0}\sin(kx)\sin(ky). \tag{S73}$$

The sign of $\psi$ alone does not determine the local flux direction. The circulation must instead be reconstructed from Eq. (S72), or equivalently inferred from the out-of-plane vorticity $(\nabla \times \mathbf{J})_z$ .

A local expansion about an isolated extremum illustrates this distinction. For a maximum at the origin,

$$\psi(x,y) \simeq \psi_0 - \frac{\kappa}{2}(x^2 + y^2), \quad \kappa > 0, \tag{S74}$$

where $\kappa > 0$ characterizes the local curvature of the stream function at the isolated extremum. one obtains

$$J_x = -\kappa y, \quad J_y = \kappa x, \tag{S75}$$

which is counterclockwise when viewed from $+z$. For a local minimum,

$$\psi(x,y) \simeq \psi_0 + \frac{\kappa}{2}(x^2 + y^2), \tag{S76}$$

the flux becomes

$$J_x = \kappa y, \quad J_y = -\kappa x, \tag{S77}$$

which is clockwise when viewed from $+z$. The handedness is therefore set by the circulation near a stream-function extremum, not by the value of $\psi$ itself.

Within a periodic unit cell, the in-plane energy-flux streamlines form closed loops. The contour $\psi$=0 separates neighboring vortices, and the extrema of $\psi$ occur at

$$\sin(kx) = \pm 1, \qquad \sin(ky) = \pm 1, \tag{S78}$$

These positions coincide with the phase singularities. The circulation reconstructed from Eq. (S72) is therefore in one-to-one correspondence with the charge of the neighboring singularity, unifying the energy-flux and phase-singularity topologies. With the viewing convention above and the circulation reconstructed from Eq. (S72), $q$=+1 corresponds to counterclockwise phase winding and the associated right-handed (RH) circulation, while $q$=-1 corresponds to the opposite left-handed (LH) circulation. This mapping provides the basis for the acoustic-spin meron lattice.

**S5. Time dependence of instantaneous velocity textures in the orthogonal standing waves ($\varphi_1$=0 limit)**

To distinguish the stationary acoustic-spin lattice from an instantaneous velocity texture, we examine the real orthogonal standing-wave field at $\varphi_1 = 0$ . The complex pressure is then purely real and has nodal lines rather than isolated phase singularities. Nevertheless, snapshots of the time-harmonic velocity can resemble a meron-like vector texture within a local region. We define the instantaneous velocity as

$$\mathbf{u}(\mathbf{r},t) = \mathrm{Re}\left[\mathbf{u}(\mathbf{r})e^{-i\omega t}\right], \tag{S79}$$

where $\mathbf{u}(\mathbf{r})$ is the complex velocity amplitude and $\omega$ is the angular frequency.

Where the instantaneous velocity is nonzero, we define its normalized direction by

$$\mathbf{n}_v(\mathbf{r},t) = \frac{\mathbf{u}(\mathbf{r},t)}{|\mathbf{u}(\mathbf{r},t)|}, \tag{S80}$$

and use the diagnostic instantaneous skyrmion number

$$Q_v(t) = \frac{1}{4\pi}\iint \mathbf{n}_v \cdot (\partial_x \mathbf{n}_v \times \partial_y \mathbf{n}_v)dxdy, \tag{S81}$$

integrated over the region outlined by the yellow dashed frame in Fig. S2(b). The quantity $Q_v(t)$ characterizes a velocity-direction snapshot and is distinct from the stationary skyrmion number defined from the complex acoustic-spin field. It is undefined at points or instants where the instantaneous velocity vanishes.

Figure S2(a) shows representative velocity snapshots over one period. A local meron-like pattern can appear, but its polarity reverses and its spatial arrangement changes with time; at zero-velocity instants the normalized direction is undefined. Correspondingly, $Q_v(t)$ in the selected region oscillates strongly and changes sign [Fig. S2(b)]. Thus $Q_v(t)$ is a time-dependent diagnostic of transient snapshots, not a conserved topological invariant of the harmonic field.

The $\varphi_1 = 0$ instantaneous-velocity texture is therefore intrinsically time dependent. This contrasts with the acoustic-spin meron lattice, which is defined from complex amplitudes and remains stationary under harmonic evolution. The comparison motivates acoustic spin, rather than an instantaneous velocity direction, as the appropriate field for time-stationary acoustic meron textures.

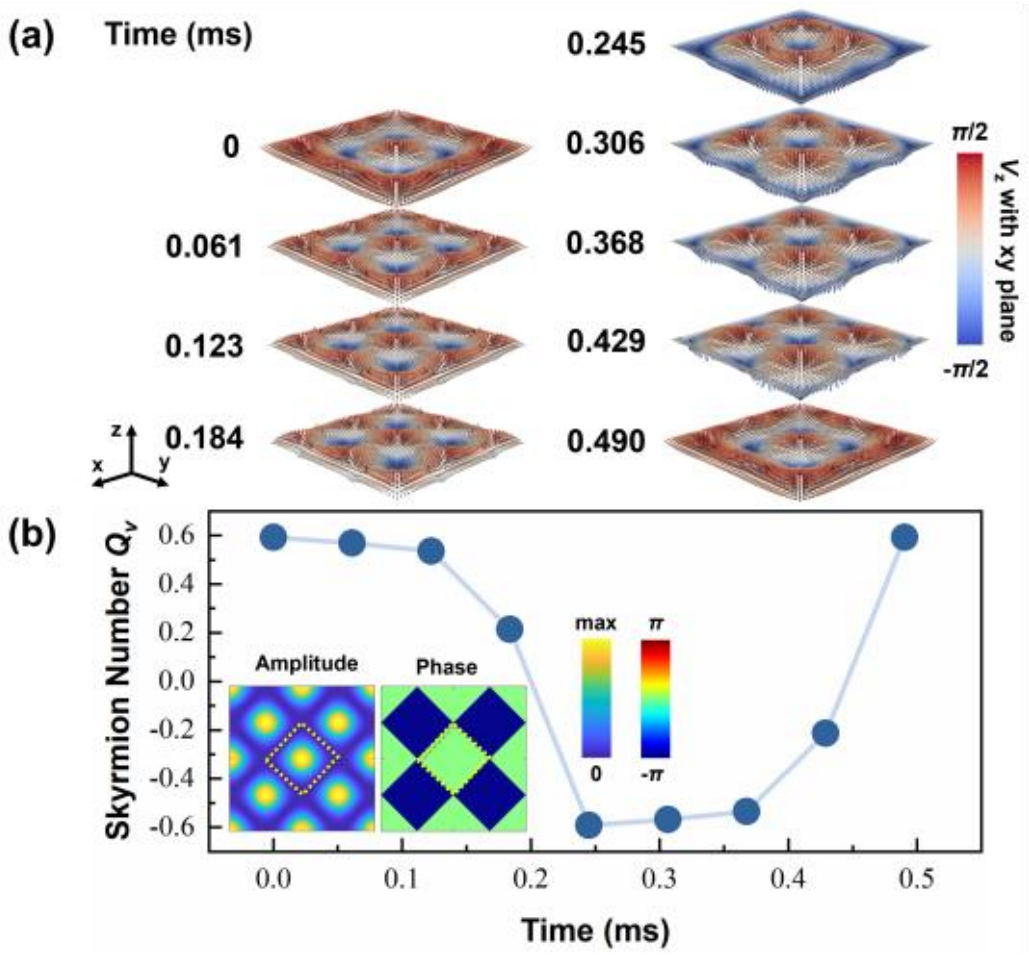

**FIG. S2. Time dependence of the instantaneous velocity texture at $\varphi_1$=0.** (a) Velocity-vector snapshots at representative times during one period; color denotes the out-of-plane component. (b) Instantaneous diagnostic $Q_v$ inset within the region marked by the yellow dashed frame. Insets show the corresponding metasurface amplitude and phase distributions.

## S6. Tunability and robustness of the acoustic-spin meron lattice

The out-of-plane component scales as $AB\sin\varphi_1$, allowing control through the excitation parameters. We sample the local texture at the position in Fig. S3(a). Simulation and experiment [Figs. S3(b) and S3(c)] show that $S_z$ varies sinusoidally with $\varphi_1$. With $B$ fixed, its magnitude increases approximately linearly with $A/B$ over the measured range. Profiles along the yellow path [Figs. S3(d) and S3(e)] show that the spin vanishes at $\varphi_1$=0, $\pi$, and 2$\pi$ and reverses polarity as these points are crossed, whereas $A/B$ mainly changes the magnitude. The interchannel phase difference and amplitude ratio therefore control, respectively, the polarity and strength of the meron lattice.

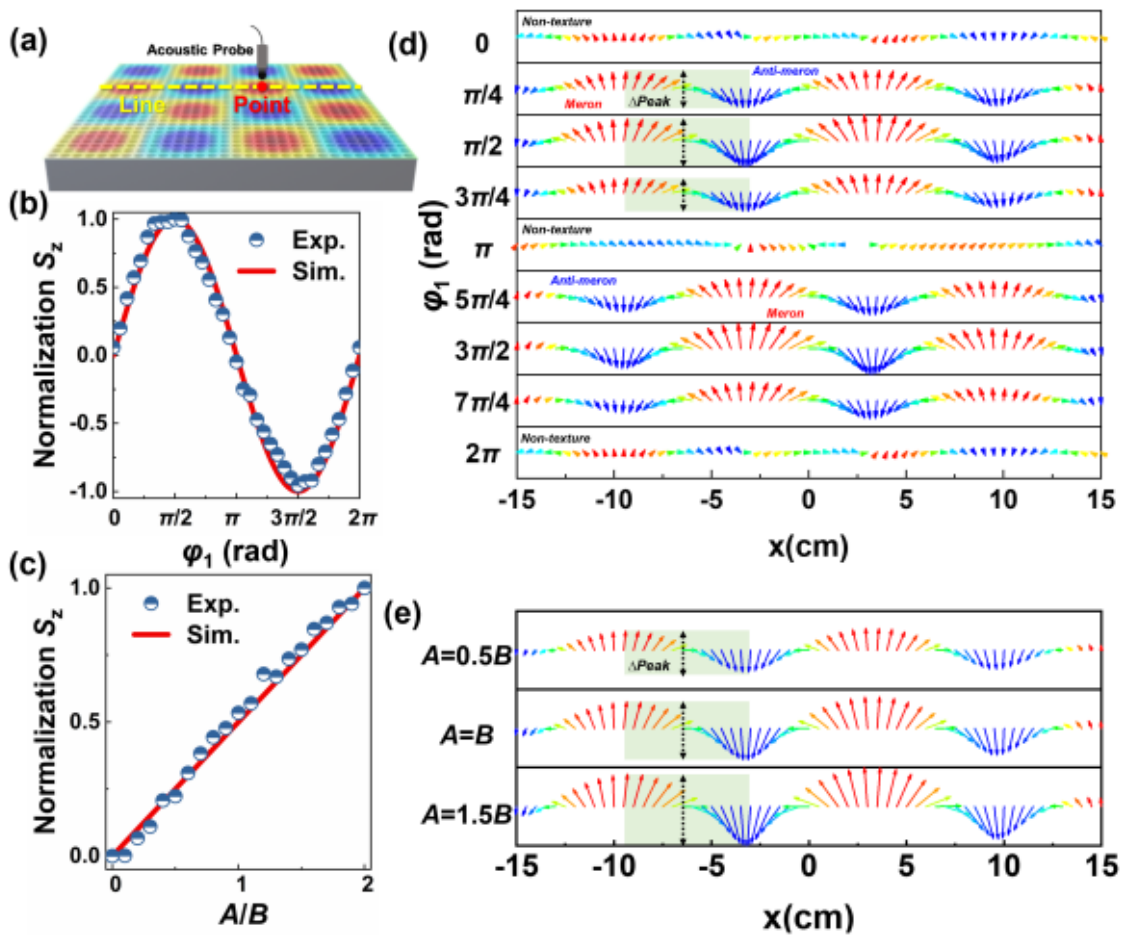

**FIG. S3. Tunability of the acoustic-spin meron lattice.** (a) Sampling position and profile path. (b), (c) Simulated and measured $S_z$ versus the interchannel phase difference $\varphi_1 = \Delta\varphi_{10}$ and amplitude ratio $A/B$, respectively. (d), (e) Spin texture profiles under variation of $\varphi_1$ and $A/B$, respectively.

We tested two structural perturbations. In the first [Fig. S4(a)], two rows of 2-cm-high scatterers were placed along opposite boundaries of the test region to disturb propagation. In the second [Fig. S4(d)], a 2×6 array of holes was sealed, locally modifying the metasurface dispersion. Simulated and measured textures for both perturbations [Figs. S4(b), S4(c), S4(e), and S4(f)] retain an identifiable 2×2 meron-lattice pattern.

Integrating the experimental skyrmion density over the 2×2 region gives $Q$=0.002 and 0.027, both close to zero and consistent with cancellation in an alternating meron/anti-meron array. Within the perturbations tested, the lattice therefore remains identifiable. Sealing holes slightly shifts $k_{ssaw}$ and perturbs the central texture more strongly than the boundary scatterers.

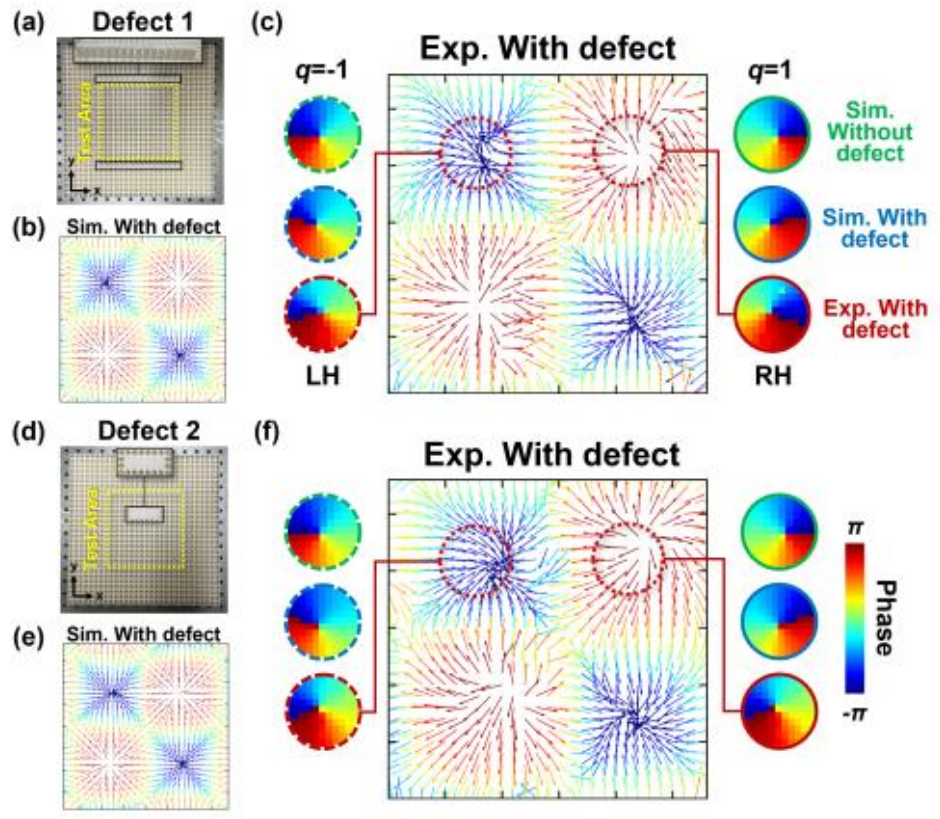

**FIG. S4. Response of the acoustic-spin meron lattice to structural defects.** (a) Boundary-scatterer configuration. (b), (c) Simulated and measured textures with boundary scatterers. (d) Sealed-hole configuration. (e), (f) Simulated and measured textures with sealed holes.

## S7. $D_6$ and $D_8$ phase encoding and pairwise-channel decomposition

The main text selects specific phase encodings from the $D_6$ and $D_8$ parameter spaces to generate the double-skyrmion superlattice and the eightfold acoustic-spin quasicrystal. Here we supplement these results by showing representative field responses for other phase combinations and by explicitly resolving the selected states into their sum- and difference-wave-vector components. The additional phase combinations are included to illustrate the field variation within the corresponding parameter spaces and are not discussed individually.

Throughout this section, the sum and difference wave vectors follow the definitions in Sec. S2,

$$\mathbf{K}_{jl}^{+} = k(\hat{\mathbf{n}}_j + \hat{\mathbf{n}}_l), \qquad \mathbf{K}_{jl}^{-} = k(\hat{\mathbf{n}}_j - \hat{\mathbf{n}}_l), \tag{S82}$$

Here, $\mathbf{K}_{jl}^{+}$ and $\mathbf{K}_{jl}^{-}$ denote the individual sum and difference wave vectors of pairwise channel $(j,l)$, whereas $K_<$ and $K_>$ denote the lower- and higher-magnitude reciprocal-space shells, respectively.

### A. $D_6$ phase encoding and two-shell decomposition

The $D_6$ configuration contains three independent standing-wave channel orientations,

$$\Theta_j = \frac{j\pi}{3}, \qquad j = 0,1,2. \tag{S83}$$

Choosing channel 0 as the phase referenc $\varphi_0 = 0$, the complex pressure at the center is $p_3(0,0) = 2A\left(1 + e^{i\varphi_1} + e^{i\varphi_2}\right)$.The condition $p_3(0,0) = 0$ gives two isolated zeros in the two-dimensional $(\varphi_1, \varphi_2)$ parameter space, $(120^\circ, 240^\circ)$ and $(240^\circ, 120^\circ)$.The first encoding is used in the main text and generates the central second-order phase singularity with $q = +2$, whereas the second produces the conjugate core with $q = -2$. The two pressure fields are related by complex conjugation, $p_3^{(240^\circ,120^\circ)}(\mathbf{r}) = \left[p_3^{(120^\circ,240^\circ)}(\mathbf{r})\right]^*$. Accordingly, $\mathbf{J}_3^{(240^\circ,120^\circ)} = -\mathbf{J}_3^{(120^\circ,240^\circ)}$, and $\mathbf{S}_3^{(240^\circ,120^\circ)} = -\mathbf{S}_3^{(120^\circ,240^\circ)}$.Thus, complex conjugation reverses the phase winding, energy-flux circulation, and acoustic-spin polarity without requiring an additional spatial rotation or mirror transformation.

Figure S5 additionally displays the pressure amplitude, phase, and acoustic-spin responses for the other discrete phase combinations considered in the two-dimensional parameter space. These additional states are shown to illustrate the field variation away from the two selected zero-pressure points and are not further classified here.

For the phase encoding selected in the main text,

$$\Theta_j = \frac{j\pi}{3}, \qquad \varphi_j = \frac{2j\pi}{3}, \qquad j = 0,1,2. \tag{S84}$$

The three pairwise interference channels are (0,1), (1,2), and (0,2). The first two have $s$=1, whereas the last has $s$=2. From the general relation in Sec. S2,

$$s = 1: \qquad |\mathbf{K}^-| = k, \qquad |\mathbf{K}^+| = \sqrt{3},k, \tag{S85}$$

whereas

$$s = 2: \qquad |\mathbf{K}^-| = \sqrt{3},k, \qquad |\mathbf{K}^+| = k. \tag{S86}$$

The sum- and difference-wave-vector harmonics therefore reorganize into two reciprocal-space shells. The lower-magnitude shell is

$$\left\{\mathbf{K}_{01}^-, \mathbf{K}_{12}^-, \mathbf{K}_{02}^+\right\}, \qquad |\mathbf{K}| = K_< = k, \tag{S87}$$

whereas the higher-magnitude shell is

$$\left\{\mathbf{K}_{01}^+, \mathbf{K}_{12}^+, \mathbf{K}_{02}^-\right\}, \qquad |\mathbf{K}| = K_> = \sqrt{3}k. \tag{S88}$$

The in-plane energy flux can consequently be decomposed as

$$\mathbf{J}_{3,\|} = \mathbf{J}_{3,\|}^{(<)} + \mathbf{J}_{3,\|}^{(>)},$$

$$\mathbf{J}_{3,\|}^{(<)} = \begin{pmatrix} \dfrac{\sqrt{3}A^2 \mathrm{e}^{-2z\tau}k}{2\rho\omega}\left(3\sin\left[\dfrac{kx}{2}-\dfrac{1}{2}\sqrt{3}ky\right]-3\sin\left[\dfrac{kx}{2}+\dfrac{1}{2}\sqrt{3}ky\right]\right) \\ \dfrac{3A^2 \mathrm{e}^{-2z\tau}k}{\rho\omega}\left(2\sin\left[kx\right]+\sin\left[\dfrac{kx}{2}-\dfrac{1}{2}\sqrt{3}ky\right]+\sin\left[\dfrac{kx}{2}+\dfrac{1}{2}\sqrt{3}ky\right]\right) \end{pmatrix},$$

$$\mathbf{J}_{3,\|}^{(>)} = \begin{pmatrix} \dfrac{\sqrt{3}A^2 \mathrm{e}^{-2z\tau}k}{2\rho\omega}\left(2\sin\left[\sqrt{3}ky\right]-\sin\left[\dfrac{3kx}{2}-\dfrac{1}{2}\sqrt{3}ky\right]+\sin\left[\dfrac{3kx}{2}+\dfrac{1}{2}\sqrt{3}ky\right]\right) \\ \dfrac{3A^2 \mathrm{e}^{-2z\tau}k}{\rho\omega}\left(-\sin\left[\dfrac{3kx}{2}-\dfrac{1}{2}\sqrt{3}ky\right]-\sin\left[\dfrac{3kx}{2}+\dfrac{1}{2}\sqrt{3}ky\right]\right) \end{pmatrix}. \tag{S89}$$

The acoustic spin component therefore satisfies

$$\mathbf{S}_3 = \mathbf{S}_3^{(<)} + \mathbf{S}_3^{(>)},$$

$$\mathbf{S}_3^{(<)} = \begin{pmatrix} \dfrac{3A^2 \mathrm{e}^{-2z\tau}k\tau}{2\rho\omega^3}\left(2\sin\left[kx\right]+\sin\left[\dfrac{kx}{2}-\dfrac{1}{2}\sqrt{3}ky\right]-\sin\left[\dfrac{3kx}{2}+\dfrac{1}{2}\sqrt{3}ky\right]\right) \\ -\dfrac{\sqrt{3}A^2 \mathrm{e}^{-2z\tau}k\tau}{2\rho\omega^3}\left(3\sin\left[\dfrac{kx}{2}-\dfrac{1}{2}\sqrt{3}ky\right]-3\sin\left[\dfrac{kx}{2}+\dfrac{1}{2}\sqrt{3}ky\right]\right) \\ \dfrac{3A^2 \mathrm{e}^{-2z\tau}k^2}{2\rho\omega^3}\left(\cos\left[kx\right]+\cos\left[\dfrac{kx}{2}-\dfrac{1}{2}\sqrt{3}ky\right]+\cos\left[\dfrac{kx}{2}+\dfrac{1}{2}\sqrt{3}ky\right]\right) \end{pmatrix},$$

$$\mathbf{S}_3^{(>)} = \begin{pmatrix} \dfrac{3A^2 \mathrm{e}^{-2z\tau}k\tau}{2\rho\omega^3}\left(-\sin\left[\dfrac{3kx}{2}-\dfrac{1}{2}\sqrt{3}ky\right]-\sin\left[\dfrac{3kx}{2}+\dfrac{1}{2}\sqrt{3}ky\right]\right) \\ -\dfrac{\sqrt{3}A^2 \mathrm{e}^{-2z\tau}k\tau}{2\rho\omega^3}\left(2\sin\left[\sqrt{3}ky\right]-\sin\left[\dfrac{3kx}{2}-\dfrac{1}{2}\sqrt{3}ky\right]+\sin\left[\dfrac{3kx}{2}+\dfrac{1}{2}\sqrt{3}ky\right]\right) \\ \dfrac{3A^2 \mathrm{e}^{-2z\tau}k^2}{2\rho\omega^3}\left(-\cos\left[\sqrt{3}ky\right]-\cos\left[\dfrac{3kx}{2}-\dfrac{1}{2}\sqrt{3}ky\right]-\cos\left[\dfrac{3kx}{2}+\dfrac{1}{2}\sqrt{3}ky\right]\right) \end{pmatrix}. \tag{S90}$$

Each reciprocal-space shell therefore contains three equal-magnitude spatial harmonics and generates a spin subtexture with a distinct spatial scale. These two shell-resolved textures correspond to the two interpenetrating skyrmion sublattices identified in the main text, and their superposition produces the $D_6$ double-skyrmion superlattice.

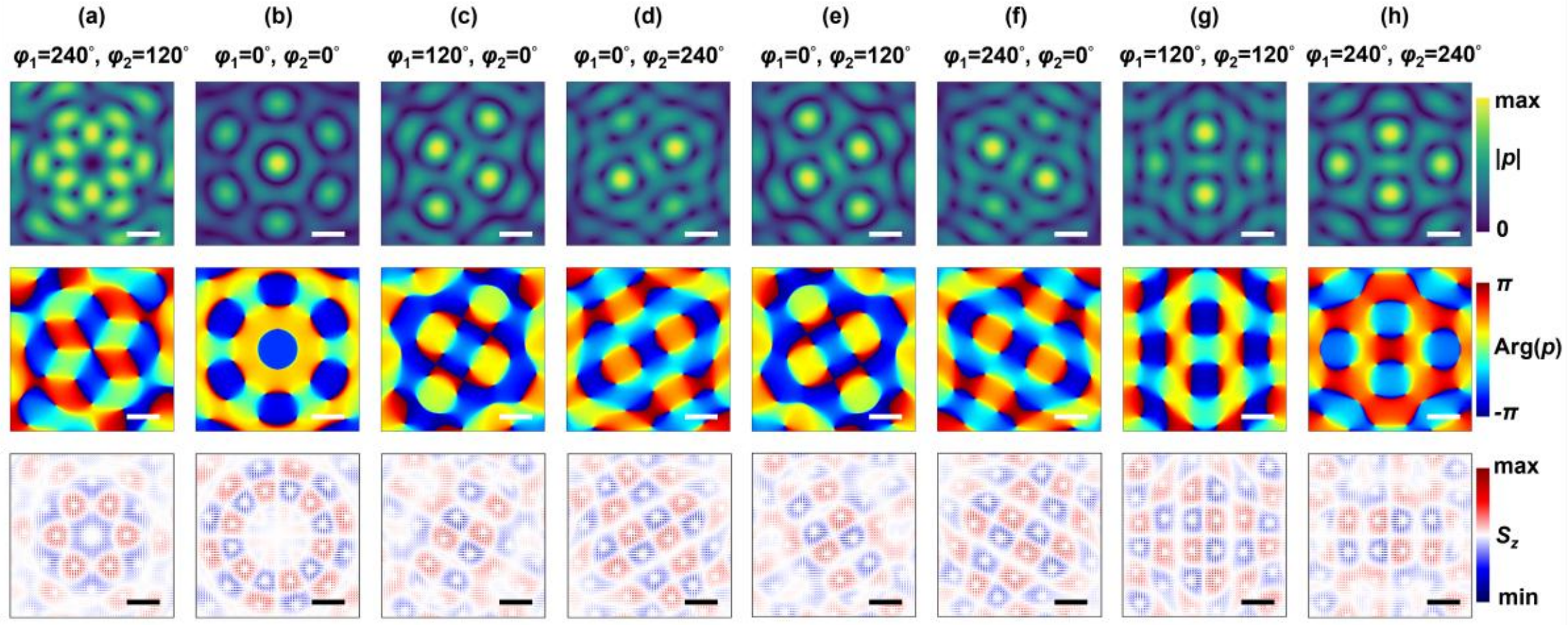

**FIG. S5. Phase dependence of the $D_6$ standing-wave field.** Representative pressure amplitude, phase, and acoustic-spin textures for discrete phase combinations in the $(\varphi_1, \varphi_2)$ parameter space.

### B. $D_8$ phase encoding and two-shell decomposition

For the equal-amplitude $D_8$ field, the four independent standing-wave channel orientations are

$$\Theta_j = \frac{j\pi}{4}, \qquad j = 0,1,2,3. \tag{S91}$$

For equal amplitudes, choosing channel 0 as the phase reference $\varphi_0 = 0$, the complex acoustic pressure at the center is

$$p_4(0,z) = 2Ae^{-\tau z}\left(1 + e^{i\varphi_1} + e^{i\varphi_2} + e^{i\varphi_3}\right). \tag{S92}$$

Unlike the two isolated zeros in the $D_6$ parameter space, the central-pressure-zero condition $p_4\left(0,z\right) = 0$ forms three sets of singular lines in the three-dimensional phase-parameter space $(\varphi_1, \varphi_2, \varphi_3)$ :

$$\begin{aligned} L_1: &\quad \varphi_1 = \pi, \quad \varphi_3 = \varphi_2 + \pi, \\ L_2: &\quad \varphi_2 = \pi, \quad \varphi_3 = \varphi_1 + \pi, \\ L_3: &\quad \varphi_3 = \pi, \quad \varphi_2 = \varphi_1 + \pi. \end{aligned} \tag{S93}$$

Figs. S6–S8 show the acoustic-pressure amplitude, acoustic-pressure phase, and acoustic-spin texture, respectively, sampled along the three singular lines, illustrating the evolution of the field response along the central-pressure-zero manifold. In the main text, we choose the phase combination $(\varphi_1, \varphi_2, \varphi_3) = \left(\frac{\pi}{2}, \pi, \frac{3\pi}{2}\right)$, whose complex-conjugate phase combination is $\left(\frac{3\pi}{2}, \pi, \frac{\pi}{2}\right)$.The two states produce opposite phase windings, energy-flux circulations, and acoustic-spin polarities. For the state selected in the main text, the six possible pairwise-interference channels are $(0,1),(1,2),(2,3),(0,3),(0,2)$, and $(1,3)$. According to the phase factor $\sin\Delta\varphi_{jl}$ introduced in Sec. S2,

$$\begin{aligned} &\sin\Delta\varphi_{01} = \sin\Delta\varphi_{12} = \sin\Delta\varphi_{23} = 1, \\ &\sin\Delta\varphi_{03} = -1, \\ &\sin\Delta\varphi_{02} = \sin\Delta\varphi_{13} = 0. \end{aligned} \tag{S94}$$

Therefore, the channels that contribute to the time-averaged energy flux and acoustic spin are $(0,1),(1,2),(2,3)$ and $(0,3)$, whereas $(0,2)$ and $(1,3)$ produce no pairwise-interference contribution. For the first three pairs of neighboring channel orientations, $\Delta\Theta = \frac{\pi}{4}$, whereas for the $(0,3)$ channel, $\Delta\Theta = \frac{3\pi}{4}$. According to the definitions of the sum and difference wave vectors in Sec. S2, the magnitudes for neighboring channels are

$$\begin{aligned} \left|\mathbf{K}_{jl}^{-}\right| &= 2k\sin\frac{\pi}{8}, \\ \left|\mathbf{K}_{jl}^{+}\right| &= 2k\cos\frac{\pi}{8}. \end{aligned} \tag{S95}$$

For the $(0,3)$ channel, the magnitudes of the sum and difference wave vectors are interchanged. Consequently, all nonzero spatial harmonics are distributed over only two reciprocal-space shells. The shell with the smaller wave-vector magnitude is $K_{<} = 2k\sin\frac{\pi}{8}$, and is composed of $\left\{\mathbf{K}_{01}^{-}, \mathbf{K}_{12}^{-}, \mathbf{K}_{23}^{-}, \mathbf{K}_{03}^{+}\right\}$. The shell with the larger wave-vector magnitude is $K_{>} = 2k\cos\frac{\pi}{8}$, and is composed of $\left\{\mathbf{K}_{01}^{+}, \mathbf{K}_{12}^{+}, \mathbf{K}_{23}^{+}, \mathbf{K}_{03}^{-}\right\}$.

Unlike the $D_6$ case, in which each shell contains three equivalent spatial-harmonic directions, each shell of the $D_8$ field contains four wave-vector directions and can be further divided into two orthogonal harmonic subsets rotated by $\pi/4$ with respect to each other. The acoustic spin can therefore be decomposed into four groups of contributions:

$$\mathbf{S}_4=\mathbf{S}_{4A}^{(<)}+\mathbf{S}_{4B}^{(<)}+\mathbf{S}_{4A}^{(>)}+\mathbf{S}_{4B}^{(>)},$$

$$\mathbf{S}_4=\begin{pmatrix}\dfrac{4A^2\mathrm{e}^{-2z\tau}k\tau}{\rho\omega^3}\sin\left(\dfrac{kx}{\sqrt{2}}\right)\left[\sqrt{2}\left(-\cos(kx)+\cos(ky)\right)\cos\left(\dfrac{ky}{\sqrt{2}}\right)+2\sin(ky)\sin\left(\dfrac{ky}{\sqrt{2}}\right)\right]\\ \dfrac{4A^2\mathrm{e}^{-2z\tau}k\tau}{\rho\omega^3}\sin\left(\dfrac{ky}{\sqrt{2}}\right)\left\{\sqrt{2}\cos\left(\dfrac{kx}{\sqrt{2}}\right)\left[\cos(kx)-\cos(ky)\right]+2\sin(kx)\sin\left(\dfrac{kx}{\sqrt{2}}\right)\right\}\\ \dfrac{4\sqrt{2}A^2\mathrm{e}^{-2z\tau}k^2}{\rho\omega^3}\left[\cos\left(\dfrac{ky}{\sqrt{2}}\right)\sin(kx)\sin\left(\dfrac{kx}{\sqrt{2}}\right)+\cos\left(\dfrac{kx}{\sqrt{2}}\right)\sin(ky)\sin\left(\dfrac{ky}{\sqrt{2}}\right)\right]\end{pmatrix}. \tag{S96}$$

The two subtextures associated with the smaller-wave-vector shell are

$$\mathbf{S}_{4A}^{(<)}=\mathbf{S}_{01}^{-}+\mathbf{S}_{23}^{-},$$

$$\mathbf{S}_{4A}^{(<)}=\begin{pmatrix}\dfrac{2A^2k\tau e^{-2\tau z}}{\rho\omega^3}\left[\dfrac{1}{\sqrt{2}}\sin\left\{k\left[\left(1-\dfrac{1}{\sqrt{2}}\right)x-\dfrac{y}{\sqrt{2}}\right]\right\}+\left(1+\dfrac{1}{\sqrt{2}}\right)\sin\left\{k\left[\dfrac{x}{\sqrt{2}}+\left(1-\dfrac{1}{\sqrt{2}}\right)y\right]\right\}\right]\\ \dfrac{2A^2k\tau e^{-2\tau z}}{\rho\omega^3}\left[-\left(1+\dfrac{1}{\sqrt{2}}\right)\sin\left\{k\left[\left(1-\dfrac{1}{\sqrt{2}}\right)x-\dfrac{y}{\sqrt{2}}\right]\right\}+\dfrac{1}{\sqrt{2}}\sin\left\{k\left[\dfrac{x}{\sqrt{2}}+\left(1-\dfrac{1}{\sqrt{2}}\right)y\right]\right\}\right]\\ \dfrac{\sqrt{2}A^2k^2e^{-2\tau z}}{\rho\omega^3}\left[\cos\left\{k\left[\left(1-\dfrac{1}{\sqrt{2}}\right)x-\dfrac{y}{\sqrt{2}}\right]\right\}+\cos\left\{k\left[\dfrac{x}{\sqrt{2}}+\left(1-\dfrac{1}{\sqrt{2}}\right)y\right]\right\}\right]\end{pmatrix}.$$

$$\mathbf{S}_{4B}^{(<)}=\mathbf{S}_{12}^{-}+\mathbf{S}_{03}^{+}, \tag{S97}$$

$$\mathbf{S}_{4B}^{(<)}=\begin{pmatrix}\dfrac{2A^2k\tau e^{-2\tau z}}{\rho\omega^3}\left[\left(1+\dfrac{1}{\sqrt{2}}\right)\sin\left\{k\left[\dfrac{x}{\sqrt{2}}-\left(1-\dfrac{1}{\sqrt{2}}\right)y\right]\right\}+\dfrac{1}{\sqrt{2}}\sin\left\{k\left[\left(1-\dfrac{1}{\sqrt{2}}\right)x+\dfrac{y}{\sqrt{2}}\right]\right\}\right]\\ \dfrac{2A^2k\tau e^{-2\tau z}}{\rho\omega^3}\left[-\dfrac{1}{\sqrt{2}}\sin\left\{k\left[\dfrac{x}{\sqrt{2}}-\left(1-\dfrac{1}{\sqrt{2}}\right)y\right]\right\}+\left(1+\dfrac{1}{\sqrt{2}}\right)\sin\left\{k\left[\left(1-\dfrac{1}{\sqrt{2}}\right)x+\dfrac{y}{\sqrt{2}}\right]\right\}\right]\\ \dfrac{\sqrt{2}A^2k^2e^{-2\tau z}}{\rho\omega^3}\left[\cos\left\{k\left[\dfrac{x}{\sqrt{2}}-\left(1-\dfrac{1}{\sqrt{2}}\right)y\right]\right\}+\cos\left\{k\left[\left(1-\dfrac{1}{\sqrt{2}}\right)x+\dfrac{y}{\sqrt{2}}\right]\right\}\right]\end{pmatrix}.$$

The two subtextures associated with the larger-wave-vector shell are

$$\mathbf{S}_{4A}^{(>)}=\mathbf{S}_{01}^{+}+\mathbf{S}_{23}^{+},$$

$$\mathbf{S}_{4A}^{(>)}=\begin{pmatrix}\dfrac{2A^2k\tau e^{-2\tau z}}{\rho\omega^3}\left[-\dfrac{1}{\sqrt{2}}\sin\left\{k\left[\left(1+\dfrac{1}{\sqrt{2}}\right)x+\dfrac{y}{\sqrt{2}}\right]\right\}+\left(1-\dfrac{1}{\sqrt{2}}\right)\sin\left\{k\left[-\dfrac{x}{\sqrt{2}}+\left(1+\dfrac{1}{\sqrt{2}}\right)y\right]\right\}\right]\\ \dfrac{2A^2k\tau e^{-2\tau z}}{\rho\omega^3}\left[-\left(1-\dfrac{1}{\sqrt{2}}\right)\sin\left\{k\left[\left(1+\dfrac{1}{\sqrt{2}}\right)x+\dfrac{y}{\sqrt{2}}\right]\right\}-\dfrac{1}{\sqrt{2}}\sin\left\{k\left[-\dfrac{x}{\sqrt{2}}+\left(1+\dfrac{1}{\sqrt{2}}\right)y\right]\right\}\right]\\ -\dfrac{\sqrt{2}A^2k^2e^{-2\tau z}}{\rho\omega^3}\left[\cos\left\{k\left[\left(1+\dfrac{1}{\sqrt{2}}\right)x+\dfrac{y}{\sqrt{2}}\right]\right\}+\cos\left\{k\left[-\dfrac{x}{\sqrt{2}}+\left(1+\dfrac{1}{\sqrt{2}}\right)y\right]\right\}\right]\end{pmatrix}.$$

$$\mathbf{S}_{4B}^{(>)}=\mathbf{S}_{12,z}^{+}+\mathbf{S}_{03}^{-}, \tag{S98}$$

$$\mathbf{S}_{4B}^{(>)}=\begin{pmatrix}\dfrac{2A^2k\tau e^{-2\tau z}}{\rho\omega^3}\left[-\left(1-\dfrac{1}{\sqrt{2}}\right)\sin\left\{k\left[\dfrac{x}{\sqrt{2}}+\left(1+\dfrac{1}{\sqrt{2}}\right)y\right]\right\}-\dfrac{1}{\sqrt{2}}\sin\left\{k\left[\left(1+\dfrac{1}{\sqrt{2}}\right)x-\dfrac{y}{\sqrt{2}}\right]\right\}\right]\\ \dfrac{2A^2k\tau e^{-2\tau z}}{\rho\omega^3}\left[-\dfrac{1}{\sqrt{2}}\sin\left\{k\left[\dfrac{x}{\sqrt{2}}+\left(1+\dfrac{1}{\sqrt{2}}\right)y\right]\right\}+\left(1-\dfrac{1}{\sqrt{2}}\right)\sin\left\{k\left[\left(1+\dfrac{1}{\sqrt{2}}\right)x-\dfrac{y}{\sqrt{2}}\right]\right\}\right]\\ -\dfrac{\sqrt{2}A^2k^2e^{-2\tau z}}{\rho\omega^3}\left[\cos\left\{k\left[\dfrac{x}{\sqrt{2}}+\left(1+\dfrac{1}{\sqrt{2}}\right)y\right]\right\}+\cos\left\{k\left[\left(1+\dfrac{1}{\sqrt{2}}\right)x-\dfrac{y}{\sqrt{2}}\right]\right\}\right]\end{pmatrix}.$$

Here, the superscripts < and > denote the reciprocal-space shells with the smaller and larger wave-vector magnitudes, respectively, whereas $A$ and $B$ denote the two orthogonal harmonic subsets within the same shell that are mutually rotated by $\pi/4$. Specifically, $\mathbf{S}_{4A}^{(<)}$ and $\mathbf{S}_{4B}^{(<)}$ correspond to two meron/anti-meron sublattices with larger spatial periods generated by the $K_<$ shell, which are mutually rotated by $\pi/4$, whereas $\mathbf{S}_{4A}^{(>)}$ and $\mathbf{S}_{4B}^{(>)}$ correspond to two meron/anti-meron sublattices with shorter spatial periods generated by the $K_>$ shell. The $D_8$ field can therefore be understood as the superposition of four distinguishable meron/anti-meron sublattices distributed over two reciprocal-space shells [Fig. S9].

The superposition of the four sublattices produces a total acoustic-spin field with eightfold rotational order but without a simple two-dimensional translational periodicity. This shows that the $D_8$ acoustic-spin quasicrystal and the $D_6$ double-skyrmion superlattice follow the same pairwise sum- and difference-wave-vector interference mechanism. In the $D_6$ field, the spatial harmonics are organized into two sublattices with different characteristic length scales, whereas in the $D_8$ field, each reciprocal-space shell further splits into two mutually rotated sublattices, thereby producing a four-sublattice, eightfold quasiperiodic topological texture.

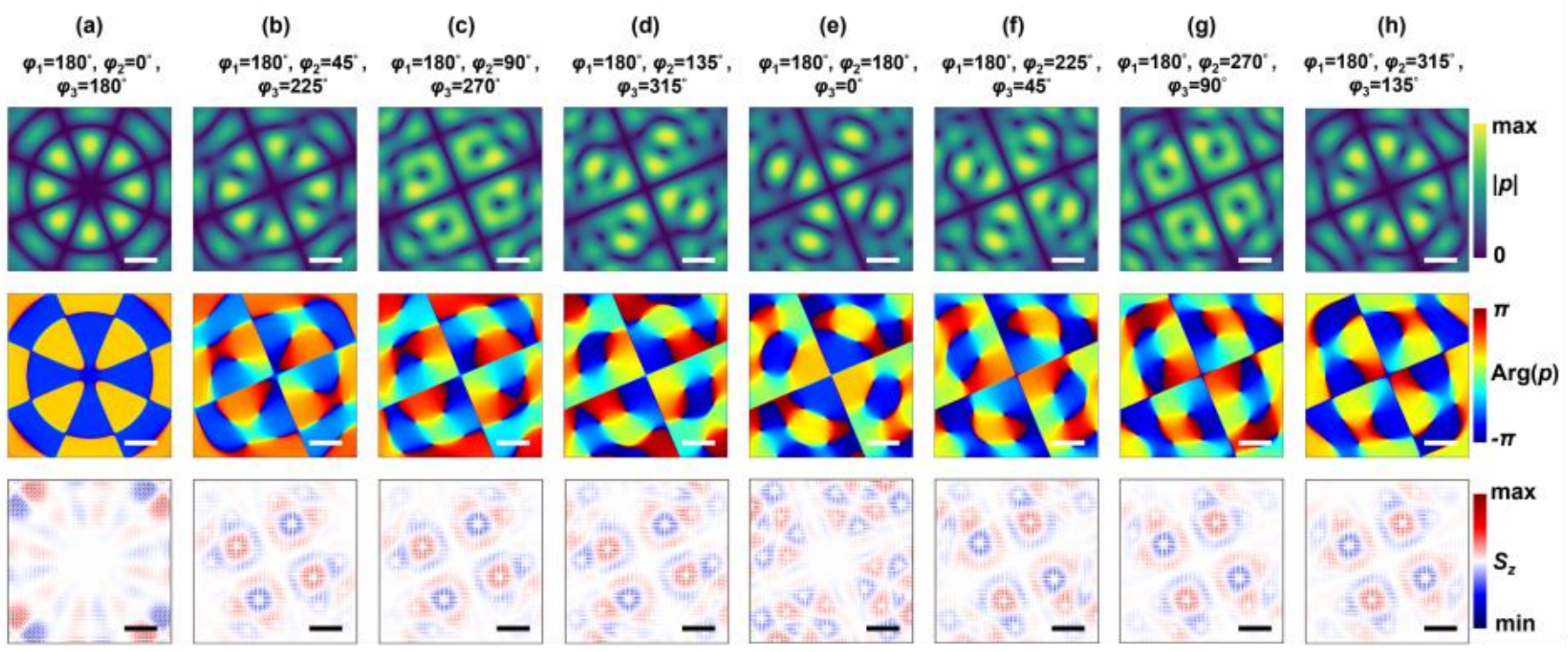

**FIG. S6. ($D_8$) field responses sampled along singular line ($L_1$).** Acoustic pressure amplitude (top), phase (middle), and acoustic-spin texture (bottom) for ($\varphi_1 = 180^\circ$) and ($\varphi_3 = \varphi_2 + 180^\circ$), with ($\varphi_2$) sampled from ($\varphi_2$) to ($315^\circ$) in ($45^\circ$) increments. The phase triplet for each column is indicated above the corresponding field maps.

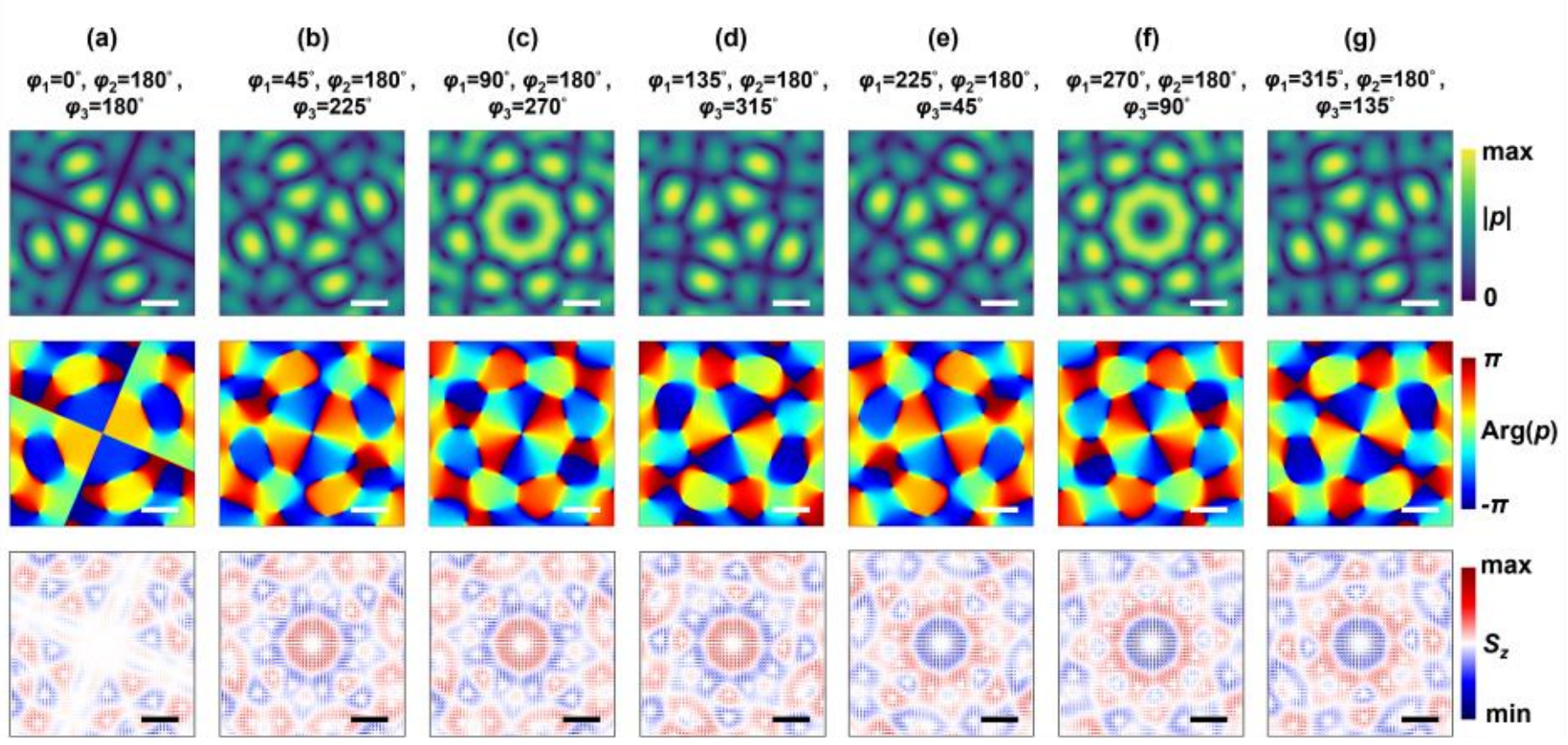

**FIG. S7. ($D_8$) field responses sampled along singular line ($L_2$).** Pressure amplitude (top), phase (middle), and acoustic-spin texture (bottom) for ($\varphi_2 = 180^\circ$) and ($\varphi_3 = \varphi_1 + 180^\circ$). The selected encoding ($90^\circ, 180^\circ, 270^\circ$) and its complex-conjugate counterpart ($270^\circ, 180^\circ, 90^\circ$) are included; the repeated intersection point is omitted.

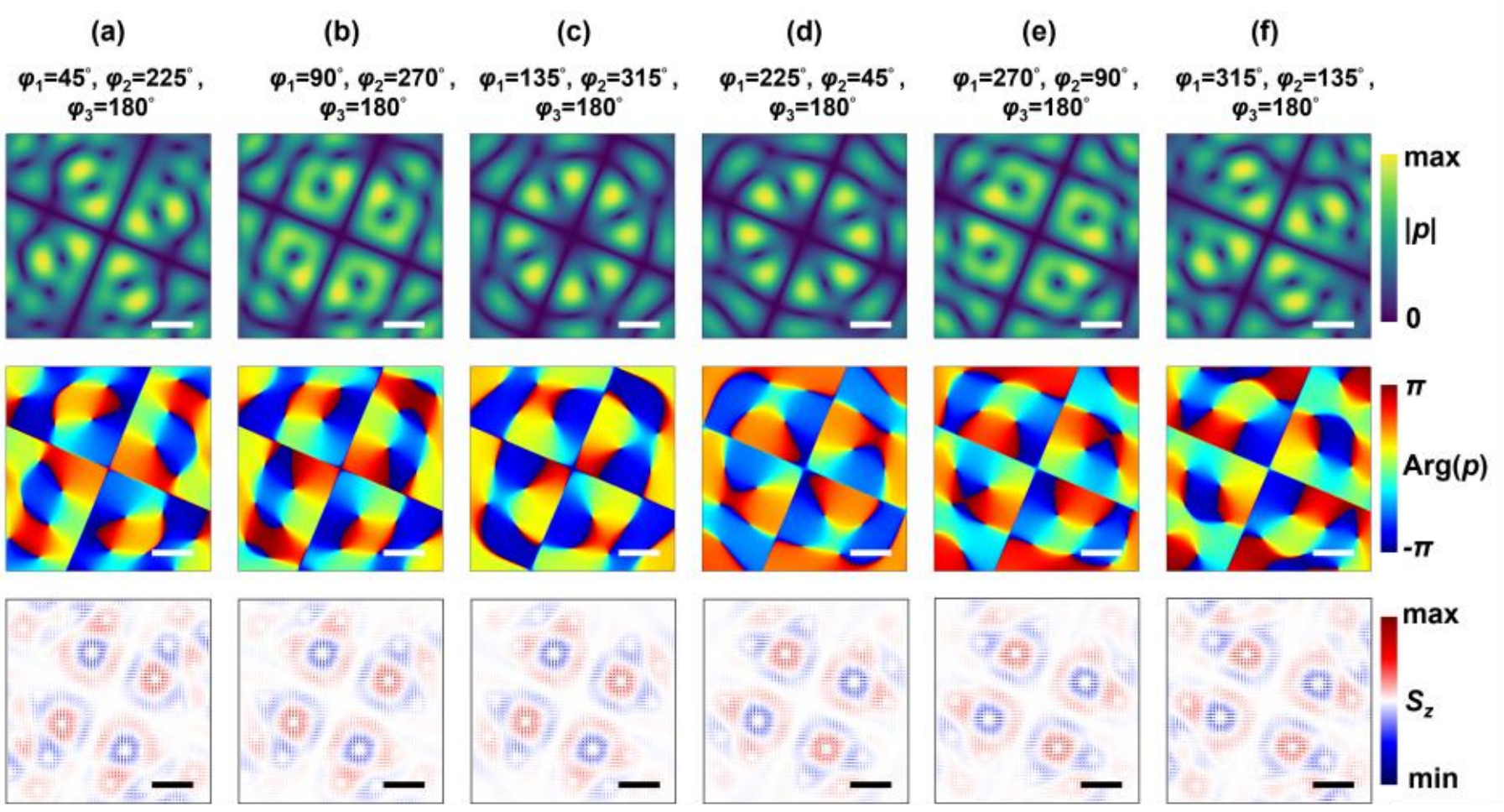

**FIG. S8. ($D_8$) field responses sampled along singular line ($L_3$).** Pressure amplitude (top), phase (middle), and acoustic-spin texture (bottom) for ($\varphi_3 = 180^\circ$) and ($\varphi_2 = \varphi_1 + 180^\circ$) . Parameter combinations already shown at intersections with ($L_1$) and ($L_2$) are omitted.

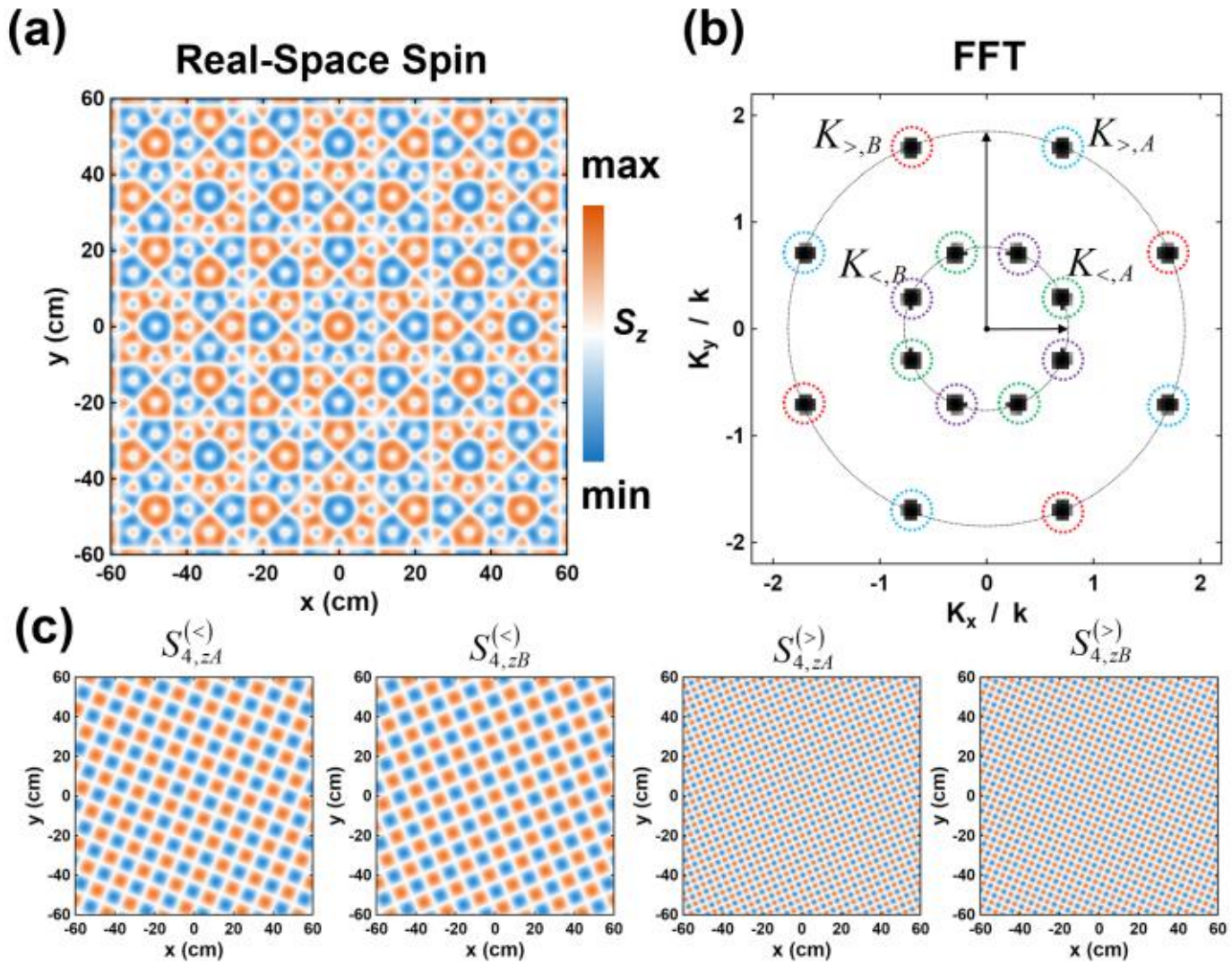

**FIG. S9. Reciprocal-space composition and four-sublattice decomposition of the $D_8$ acoustic-spin quasicrystal.** (a) Real-space distribution of the total out-of-plane acoustic spin $S_z$. (b) Two-dimensional Fourier spectrum of $S_z$. (c) Out-of-plane acoustic-spin subtextures reconstructed from the four wavevector subsets $S^{(<)}_{4,zA}, S^{(<)}_{4,zB}, S^{(>)}_{4,zA}$ and $S^{(>)}_{4,zB}$ .

**S8. Zeroth-order Bessel limit**

In the general polygonal standing-wave theory of Sec. S2, the channel-axis orientations $\Theta_j$ and the relative channel-phase encoding $\varphi_j$ constitute independent control degrees of freedom. The relative-phase combinations adopted for the $D_6$ and $D_8$ configurations in Sec. S7 are selected from zeros of the central acoustic pressure in the phase-parameter space, rather than being prescribed a priori by the standing-wave channel orientations. Here, we introduce a particular angular phase parameterization to establish the correspondence between finite polygonal standing-wave fields and continuous Bessel angular spectra.

According to the discrete standing-wave channel orientations defined in Sec. S2,

$$\Theta_j = \frac{j\pi}{N}, j = 0,1...,N-1. \tag{S99}$$

With channel 0 chosen as the phase reference, we define the angular phase encoding

$$\varphi_j = \frac{m\pi j}{N}, \tag{S100}$$

or, equivalently,

$$\varphi_j = m\Theta_j. \tag{S101}$$

Here, $m \in \mathbb{Z}$ is the discrete angular phase index. The interchannel phase difference between neighboring channels is therefore

$$\Delta\varphi = \varphi_{j+1} - \varphi_j = \frac{m\pi}{N}. \tag{S102}$$

We next consider equal amplitudes $A$ for all standing-wave channels. The acoustic-pressure field for finite $N$ [Eq.(S13)] can then be written as

$$p_N^{(m)} = 2Ae^{-\tau z}\sum_{j=0}^{N-1} e^{im\Theta_j}\cos\left(k\hat{\mathbf{n}}_j \cdot \mathbf{r}_{\|}\right). \tag{S103}$$

If the amplitude $A$ of each standing-wave channel is kept fixed, the overall pressure scale increases with $N$. To obtain a finite continuous limit and to compare the spatial field patterns for different $N$, we therefore define the dimensionless normalized field

$$\tilde{p}_N^{(m)} = \frac{p_N^{(m)}}{2ANe^{-\tau z}} = \frac{1}{N}\sum_{j=0}^{N-1} e^{im\Theta_j}\cos\left(k\hat{\mathbf{n}}_j \cdot \mathbf{r}_{\|}\right), \tag{S104}$$

Because the angular spacing is $\Delta\Theta = \pi/N$, Eq.(S104) constitutes a Riemann sum for a continuous angular integral. Therefore, as $N \to \infty$,

$$\tilde{p}_N^{(m)} \to \tilde{p}_\infty^{(m)} = \frac{1}{\pi}\int_0^\pi e^{im\Theta}\cos\left[k\hat{\mathbf{n}}(\Theta)\cdot \mathbf{r}_{\|}\right]d\Theta. \tag{S105}$$

A finite polygonal standing-wave field can therefore be interpreted as a normalized discrete sampling of a continuous standing-wave angular spectrum [Fig. S10(a)]. Under this normalization, the physical pressure field corresponding to the continuous limit is

$$p_\infty^{(m)} = 2Ae^{-\tau z}\tilde{p}_\infty^{(m)}. \tag{S106}$$

We first consider the simplest in-phase spectrum, $m = 0$. Writing the in-plane observation position in polar coordinates as $r_{\|} = r\left(\cos\alpha, \sin\alpha\right)$, we have $\hat{\mathbf{n}}\left(\Theta\right)\cdot \mathbf{r}_{\|} = r\cos\left(\Theta - \alpha\right)$.Equation (S105) therefore becomes

$$\tilde{p}_\infty^{(0)} = \frac{1}{\pi}\int_0^\pi \cos\left[kr\cos\left(\Theta - \alpha\right)\right]d\Theta. \tag{S107}$$

Letting $\beta = \Theta - \alpha$, and noting that $\cos\left(kr\cos\beta\right)$ is $\pi$ -periodic, the integration interval can be shifted without changing the integral, yielding

$$\tilde{p}_{\infty}^{(0)} = \frac{1}{\pi}\int_0^{\pi} \cos(kr\cos\beta)\,d\beta. \tag{S108}$$

Using the integral representation of the Bessel function of the first kind,

$$J_0(kr) = \frac{1}{\pi}\int_0^{\pi} \cos(kr\cos\beta)\,d\beta, \tag{S109}$$

we obtain

$$\begin{aligned} \tilde{p}_{\infty}^{(0)} &= J_0(kr), \\ p_{\infty}^{(0)} &= 2Ae^{-\tau z}J_0(kr). \end{aligned} \tag{S110}$$

Thus, under normalized angular weighting, the in-phase polygonal standing-wave angular spectrum converges to the zeroth-order Bessel standing wave as $N \to \infty$. Figure S10(b) shows this evolution in real space, the pronounced polygonal modulation at small $N$ is progressively suppressed as the angular sampling becomes denser, and the field approaches the circularly symmetric $J_0(kr)$ distribution.

To further examine this convergence, we extract radial profiles along $(y = 0, x \geq 0)$, for which $r = x$. The sampling interval is $0 \leq kr \leq 12$, as illustrated in Fig. S10(c). Figure S10(d) compares the corresponding normalized $m = 0$ radial profiles with $J_0(kr)$. For small $N$, agreement is first established near the field center, whereas appreciable deviations remain at larger $kr$. Increasing $N$ progressively extends the spatial range over which the discrete field follows the continuous Bessel profile. These results show that the finite polygonal standing-wave field can be understood as a discrete angular realization of the continuous zeroth-order Bessel spectrum, with its spatial agreement systematically improving as the angular sampling becomes denser.

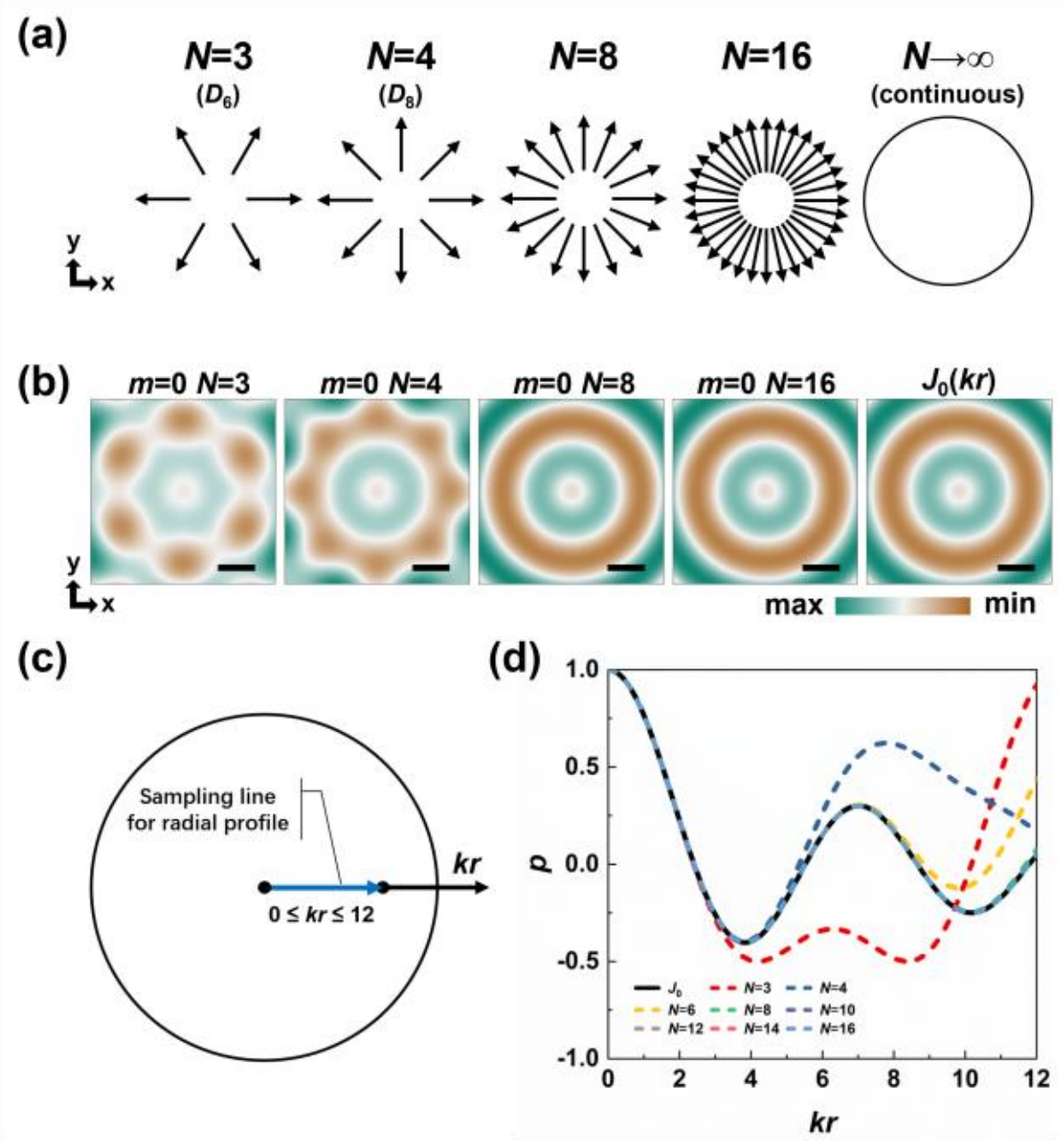

**FIG. S10. Discrete angular sampling and convergence toward the zeroth-order Bessel limit.** (a) Angular sampling for $N = 3, 4, 8$ and 16 and in the continuous limit. (b) Normalized acoustic-pressure fields for $m = 0$ with $N = 3, 4, 8$ and 16, compared with the continuous zeroth-order Bessel field $J_0(kr)$. (c) Radial sampling path along $y = 0$ over $0 \leq kr \leq 12$ used for the profiles in (d). (d) Corresponding radial profiles of the normalized $m = 0$ fields compared with $J_0(kr)$.

**S9. Second-order Bessel limit and finite angular discretization**

Section S8 considered the in-phase angular spectrum $m = 0$ . A nontrivial angular phase encoding instead produces a Bessel-vortex field with a central phase singularity. Because the standing-wave basis is invariant under $\Theta \to \Theta + \pi$ , their spatial factors satisfy

$$\cos\left[k\hat{\mathbf{n}}(\Theta+\pi)\cdot\mathbf{r}_{\parallel}\right] = \cos\left[k\hat{\mathbf{n}}(\Theta)\cdot\mathbf{r}_{\parallel}\right], \tag{S111}$$

The continuous angular weighting must therefore possess the same $\pi$ periodicity, $e^{im(\Theta+\pi)} = e^{im(\Theta)}$ . Accordingly, apart from the in-phase case $m = 0$ , the lowest-order nontrivial angular encoding allowed by the standing-wave representation is $m = 2$ . From Eq. (S105), its continuous limit is

$$\tilde{p}_{\infty}^{(2)} = \frac{1}{\pi}\int_0^{\pi} e^{i2\Theta}\cos\left[kr\cos(\Theta-\alpha)\right]d\Theta. \tag{S112}$$

Letting $\beta = \Theta - \alpha$ , we have $e^{i2\Theta} = e^{i2\alpha}e^{i2\beta}$ . Because the complete integrand $e^{i2\beta}\cos(kr\cos\beta)$ remains $\pi$ -periodic, the integration interval can again be shifted, giving

$$\tilde{p}_{\infty}^{(2)} = \frac{e^{i2\alpha}}{\pi}\int_0^{\pi} e^{i2\beta}\cos(kr\cos\beta)\,d\beta. \tag{S113}$$

Using the Bessel expansion of the cosine function,

$$\cos(x\cos\beta) = J_0(x) + 2\sum_{\ell=1}^{\infty}(-1)^{\ell}J_{2\ell}(x)\cos(2\ell\beta). \tag{S114}$$

Using Eq. (S114)the angular integral can be written as

$$\frac{1}{\pi}\int_0^{\pi} e^{i2\beta}\cos(kr\cos\beta)\,d\beta = \frac{J_0(kr)}{\pi}\int_0^{\pi} e^{i2\beta}d\beta + \frac{2}{\pi}\sum_{\ell=1}^{\infty}(-1)^{\ell}J_{2\ell}(kr)\int_0^{\pi} e^{i2\beta}\cos(2\ell\beta)d\beta, \tag{S115}$$

here $\ell = 1,2,\ldots$ is the summation index. The zeroth-order term vanishes because $\int_0^{\pi} e^{i2\beta}d\beta = 0$ . For the higher-order terms, using $\cos(2\ell\beta) = \frac{1}{2}\left(e^{i2\ell\beta} + e^{-i2\ell\beta}\right)$ , gives

$$\int_0^{\pi} e^{i2\beta}\cos(2\ell\beta)d\beta = \frac{1}{2}\int_0^{\pi}\left[e^{i2(\ell+1)\beta} + e^{i2(1-\ell)\beta}\right]d\beta. \tag{S116}$$

For $\ell \neq 1$ , both exponential terms have nonzero integer angular order and integrate to zero over $[0,\pi]$ . Therefore,

$$\int_0^{\pi} e^{i2\beta}\cos(2\ell\beta)d\beta = 0, \quad \ell \neq 1. \tag{S117}$$

For $\ell = 1$ ,

$$\int_0^{\pi} e^{i2\beta}\cos(2\beta)d\beta = \frac{\pi}{2}. \tag{S118}$$

Hence only the $\ell = 1$ term survives, and

$$\frac{1}{\pi}\int_0^{\pi} e^{i2\beta}\cos(kr\cos\beta)\,d\beta = -J_2(kr). \tag{S119}$$

We therefore obtain

$$\begin{aligned}\tilde{p}_{\infty}^{(2)} &= -J_2(kr)e^{i2\alpha},\\ p_{\infty}^{(2)} &= -2Ae^{-\tau z}J_2(kr)e^{i2\alpha}.\end{aligned} \tag{S120}$$

Thus, the $m = 2$ angular encoding of the continuous standing-wave spectrum generates a second-order Bessel vortex. The overall minus sign in Eq. (S120) corresponds only to a uniform $\pi$ phase offset and does not affect the intensity, time-averaged energy flux, acoustic spin, or topological charge. Figure S11(a) compares the acoustic-pressure amplitude and phase for $m = 2$ with $N = 5,6,8,16$ against the continuous second-order Bessel field $J_2(kr)e^{i2\alpha}$ . The finite angular samplings with $N = 3$ and $N = 4$ correspond to the $D_6$ and $D_8$ configurations discussed in the main text, respectively, and are therefore not repeated here.

Near the origin,

$$J_2(kr) = \frac{(kr)^2}{8} + O\left[(kr)^4\right], \tag{S121}$$

and therefore

$$\begin{aligned} \tilde{p}_\infty^{(2)} &= -\frac{k^2}{8}(x+iy)^2 + O\left[(kr)^4\right], \\ p_\infty^{(2)} &= -\frac{Ak^2}{4}e^{-\tau z}(x+iy)^2 + O\left[(kr)^4\right]. \end{aligned} \tag{S122}$$

Since $x + iy = re^{i\alpha}$, the local field contains $(x+iy)^2 = r^2 e^{i2\alpha}$, and its phase therefore satisfies $\arg\left[\tilde{p}_\infty^{(2)}\right] = 2\alpha + const$. Along a counterclockwise closed loop viewed from the $+z$ direction, $\alpha$ increases from 0 to $2\pi$, giving, $\Delta\arg\left[\tilde{p}_\infty^{(2)}\right] = 4\pi$ and hence the phase singularity has topological charge $q = +2$. Its complex-conjugate field satisfies $\tilde{p}_\infty^{(2)*} \propto (x-iy)^2$, and therefore carries $q = -2$. Thus, the $m = \pm 2$ encodings correspond to a pair of conjugate second-order phase singularities with opposite winding directions.

The second-order topological core does not require the $N \to \infty$ continuous limit. For any finite angular sampling $N$, we define

$$\tilde{p}_N^{(2)} = \frac{1}{N}\sum_{j=0}^{N-1} e^{i2\Theta_j}\cos\left(k\hat{\mathbf{n}}_j \cdot \mathbf{r}_\parallel\right). \tag{S123}$$

Near the origin, the cosine term can be expanded as

$$\cos\left(k\hat{\mathbf{n}}_j \cdot \mathbf{r}_\parallel\right) = 1 - \frac{k^2}{2}\left(\hat{\mathbf{n}}_j \cdot \mathbf{r}_\parallel\right)^2 + O\left[(kr)^4\right]. \tag{S124}$$

For any $N \geq 3$, $\sum_{j=0}^{N-1} e^{i2\Theta_j} = 0$. The quadratic term satisfies

$$\sum_{j=0}^{N-1} e^{i2\Theta_j}\left(\hat{\mathbf{n}}_j \cdot \mathbf{r}_\parallel\right)^2 = \frac{N}{4}(x+iy)^2. \tag{S125}$$

Substituting Eqs. (S124) and (S125) into Eq. (S123) gives

$$\tilde{p}_N^{(2)} = -\frac{k^2}{8}(x+iy)^2 + O\left[(kr)^4\right], \quad N \geq 3. \tag{S126}$$

Equation (S126) is identical to the near-origin expansion of the continuous second-order Bessel field in Eq. (S122). Therefore, for $m = 2$, any finite angular sampling with $N \geq 3$ possesses the same local second-order topological core near the origin.

For $N = 3$, $(\Theta_0, \Theta_1, \Theta_2) = (0, \pi/3, 2\pi/3)$, and the corresponding $m = 2$ phase encoding is $(\varphi_0, \varphi_1, \varphi_2) = (0, 2\pi/3, 4\pi/3)$, which is precisely the phase encoding selected for the $D_6$ configuration in the main text. For $N = 4$, $(\Theta_0, \Theta_1, \Theta_2, \Theta_3) = (0, \pi/4, \pi/2, 3\pi/4)$, and $(\varphi_0, \varphi_1, \varphi_2, \varphi_3) = (0, \pi/2, \pi, 3\pi/2)$, which coincides with the phase encoding selected for the $D_8$ configuration. Thus, although the $D_6$ and $D_8$ phase combinations were independently identified from zeros of the central pressure in their respective phase-parameter spaces, both belong to the same $m = 2$ angular-encoding family. This explains why the $D_6$ and $D_8$ fields share the same local $q = 2$ second-order phase-singularity core.

To quantify the global convergence of finite angular sampling toward the continuous Bessel limits, we compare the normalized discrete field $\tilde{p}_N^{(m)}$ with its continuous counterpart $\tilde{p}_\infty^{(m)}$ for $m = 0$ and $m = 2$ . The radial profile in Fig. S10(d) is sampled along $y = 0, 0 \le kr \le 12$ , whereas the two-dimensional relative $L_2$ error in Fig. S11(b) is evaluated over the central disk

$$\Omega = \{(x, y) | kr \le 12\}. \tag{S127}$$

For two complex fields $f$ and $g$ , we define the inner product over $\Omega$ as

$$\langle f, g \rangle_\Omega = \int_\Omega f^*(\mathbf{r}_\parallel) g(\mathbf{r}_\parallel) d^2\mathbf{r}_\parallel, \tag{S128}$$

with the corresponding $L_2$ norm

$$\|f\|_{2,\Omega} = \sqrt{\langle f, f \rangle_\Omega}. \tag{S129}$$

Although the $1/N$ normalization removes the systematic field scaling associated with increasing $N$ , a spatially uniform complex scale factor may remain between a finite discrete field and its continuous reference. To ensure that the error measures only the spatial-field mismatch, we introduce the least-squares optimal complex scaling factor

$$c_N^{(m)} = \frac{\left\langle \tilde{p}_\infty^{(m)}, \tilde{p}_N^{(m)} \right\rangle_\Omega}{\left\langle \tilde{p}_\infty^{(m)}, \tilde{p}_\infty^{(m)} \right\rangle_\Omega}. \tag{S130}$$

The relative $L_2$ convergence error is then defined as

$$\varepsilon_N^{(m)} = \frac{\left\| p_N^{(m)} - c_N^{(m)} p_\infty^{(m)} \right\|_2}{\left\| c_N^{(m)} p_\infty^{(m)} \right\|_2}, \qquad m = 0, 2. \tag{S131}$$

The factor $c_N^{(m)}$ removes only differences in overall amplitude and global phase, so that $\varepsilon_N^{(m)}$ quantifies the spatial mismatch between the finite and continuous fields within the selected two-dimensional region.

As shown in Fig. S11(c), the relative $L_2$ errors for both $m = 0$ and $m = 2$ decrease rapidly with increasing $N$ and eventually reach the numerical-precision floor. This quantitatively confirms that normalized polygonal standing-wave fields converge to their corresponding continuous Bessel angular spectra as the angular sampling becomes dense.

The $N = 3$ and $N = 4$ samplings, corresponding to the $D_6$ and $D_8$ configurations, respectively, still exhibit appreciable global deviations from the continuous second-order Bessel field over the finite region $kr \le 12$ . This does not conflict with their exact local $q = 2$ core. The quantity $\varepsilon_N^{(2)}$ measures the global field-shape agreement over the full two-dimensional disk $\Omega$ , whereas the near-origin expansion establishes local topological equivalence in the limit $r \to 0$ . Finite $D_6$ and $D_8$ fields can therefore share the same local second-order Bessel-vortex core while retaining the distinct long-range superlattice and quasiperiodic structures imposed by their finite discrete interference channels.

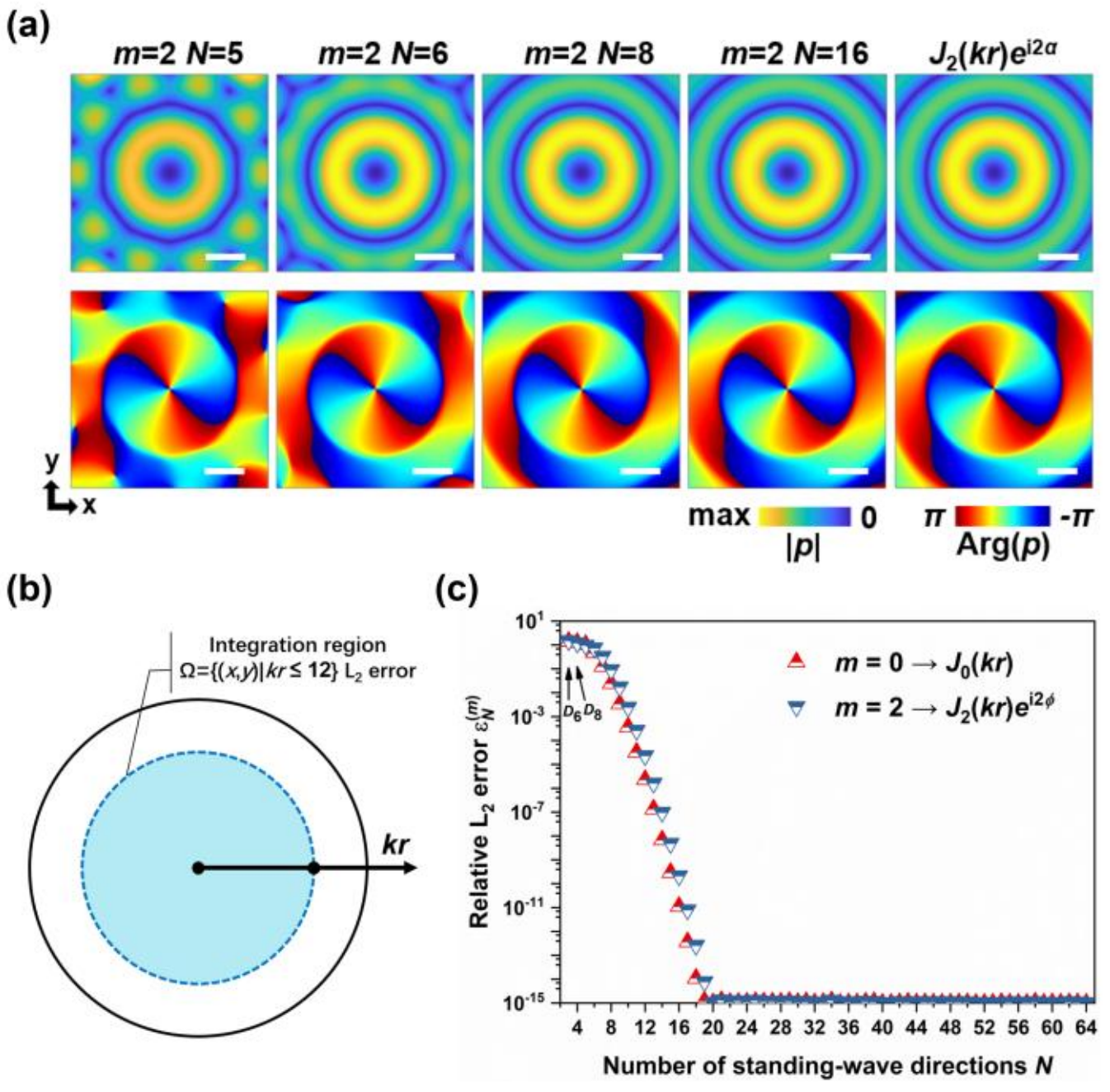

**FIG. S11. Second-order Bessel correspondence and quantitative convergence of finite angular sampling.** (a) Acoustic-pressure amplitude and phase for the $m = 2$ angular encoding with $N = 5,6,8$ and 16, compared with the continuous second-order Bessel field $J_2(kr)e^{i2\alpha}$. (b) Integration region used to evaluate the relative $L_2$ error, $\Omega = \{(x, y)|\ kr \le 12\}$. (c) Relative $L_2$ convergence errors $\varepsilon_N^{(m)}$ for the $m = 0$ and $m = 2$ angular encodings as functions of the number $N$ of standing-wave channel orientations.